%% file: PAPER_ARXIV.tex
\documentclass{article} % For LaTeX2e
\usepackage{iclr2027_conference,times}

\input{math_commands.tex}

\usepackage[utf8]{inputenc}         % allow utf-8 input
\usepackage[T1]{fontenc}            % use 8-bit T1 fonts
\usepackage{hyperref}               % hyperlinks
\usepackage{url}                    % simple URL typesetting
\usepackage{booktabs}               % professional-quality tables
\usepackage{amsfonts}               % blackboard math symbols
\usepackage{nicefrac}               % compact symbols for 1/2, etc.
\usepackage{microtype}              % microtypography
\usepackage[dvipsnames]{xcolor}     % colors
\usepackage{layouts}
\usepackage{bm}
\usepackage{amsmath}
\usepackage{graphicx}
\usepackage{physics}
\usepackage{comment}
\usepackage{wrapfig}
\usepackage{amssymb}
\usepackage{float}
\usepackage[normalem]{ulem}
\graphicspath{{figures/}}
\usepackage{algorithm}
\usepackage{algpseudocode}
\usepackage{enumitem}

\newcommand{\JL}[1]{\textcolor{blue}{[JL: #1]}}

\makeatletter
\DeclareRobustCommand{\showfontsize}{%
  \texttt{[\f@size pt, baseline \the\baselineskip]}%
}
\makeatother

\renewcommand{\eqref}[1]{(\ref{#1})}

\title{Symmetry-Aware Residual Learning of\\ Few-Body Molecular Wave Functions}

\author{%
    \makebox[\dimexpr\textwidth-2\tabcolsep\relax][c]{%
  \begin{tabular}[t]{@{}c@{}}
    Meital Bojan\thanks{These authors contributed equally to this work.},\,
    Mikhail Maslov\footnotemark[1],\,
    Jing-Lun Li\thanks{Corresponding author.}, \\
    Alex M. Bronstein, and Mikhail Lemeshko \\[0.5ex]
    Institute of Science and Technology Austria (ISTA) \\
    Am Campus 1, 3400 Klosterneuburg, Austria \\[0.5ex]
    {\small\texttt{meital.bojan@ista.ac.at},\,
    \texttt{mmaslov@ista.ac.at},\,
    \texttt{jinglun.li@ista.ac.at}}, \\
    {\small\texttt{alexander.bronstein@ista.ac.at},\,
    \texttt{mikhail.lemeshko@ista.ac.at}}
  \end{tabular}%
  }%
}

\iclrfinalcopy % Uncomment for camera-ready version, but NOT for submission.
\begin{document}

\maketitle

\begin{abstract}
Ultracold polar molecules are a highly controllable platform for engineering quantum matter, but modeling their few-body properties remains computationally challenging. Their anisotropic long-range interactions and rich internal structure lead to high-dimensional eigenvalue problems for which conventional grid-based solvers are largely restricted to two-molecule systems. To address this limitation, we introduce a physics-informed neural variational method that exploits known physical structure of the molecular wave function. Our key idea is to factor the wave function into analytic and neural components. The analytic component encodes rotational symmetry and asymptotic decay, while the neural residual learns the remaining intermediate-range behavior. This decomposition reduces the effective dimensionality of the problem without relying on ad hoc approximations or problem-specific discretization grids. We benchmark our method against high-accuracy two-molecule solvers and recover both ground and excited bound states with mean relative errors below $0.01\%$. We further demonstrate the method’s applicability to three-body systems by computing field-linked trimer ground-state energies in percent-level agreement with an independent grid-based calculation. This provides a systematic route to larger molecular systems and opens opportunities to study correlations and dynamics in interacting quantum matter.
\end{abstract}
% We also validate the numerical building blocks required for extension to three-body systems. Specifically, we test the angular-channel decomposition and two-coordinate radial representation in settings where reliable benchmarks are available.

\section{Introduction}
\label{sec_introduction}

Neural networks (NNs) have become powerful computational tools in quantum physics~\citep{Carleo_2019,Karniadakis_2021,raissi2019physics} and chemistry~\citep{Goh2017,Westermayr2021,Nomura2024,Mroz2025}. For example, they have enabled accurate descriptions of correlated spin dynamics~\citep{Carleo_2017} and molecular electronic wave functions~\citep{Pfau_2020, Hermann_2020}. A central challenge in these problems is the presence of many strongly coupled degrees of freedom, which renders conventional numerical approaches inefficient at high dimensions~\citep{Troyer_2005,Schuch_2009}. NNs provide general and flexible representations of candidate solutions that can capture nontrivial quantum structure through direct optimization of the underlying physical objective.
While many-body physics has been extensively studied in this context, the full quantum dynamics of few-particle systems with realistic interactions likewise present a high-dimensional NP-hard challenge~\citep{Troyer_2005,Kempe_2006} well suited to flexible neural representations.

Ultracold polar molecules with anisotropic long-range interactions coupling multiple internal degrees of freedom represent such a physical setting. They have emerged as a versatile experimental platform for engineered quantum matter due to their high controllability and rich internal structure~\citep{Carr2009,Moses_2017}, enabling a wide range of applications from exotic quantum phases~\citep{Baranov2012} to quantum simulation and computation \citep{DeMille_2002,Wall2015,Cornish2024}. For fully exploiting control over such systems, it is first crucial to understand the few-body properties of molecules, as they encode the effective interactions and correlations that govern their collective behavior. 
% However, due to the intricate interplay between the anisotropic interaction and molecular internal structure, solving such few-body problems with conventional grid-based methods is primarily limited to two-molecule systems. 
Solving molecular few-body problems with conventional grid-based methods is primarily limited to two- and three-molecule systems. 
%Introducing new computational tools in this nascent field offers substantial potential for uncovering emergent phenomena and deepening physical understanding.
This limitation motivates new computational approaches that can broaden the range of tractable molecular quantum problems.

In this work, we introduce a NN representation of molecular wave functions \mbox{to address this bottleneck.} A general NN without an inductive bias would have to infer from optimization all physical properties that are characteristic of molecular few-body Hamiltonians, including the overall translational symmetry, angular structure, and asymptotic decay. Instead, our \textit{central idea} is to explicitly incorporate this general physical structure into the wave function representation, leaving the network to learn a symmetry-aware residual. We realize this idea by factoring the full wave function into an analytic component that encodes known structure, and a neural residual that captures the remaining correlations. The angular dependence is represented using spherical harmonics, reducing the problem to coupled radial channels, and an exponential envelope imposes the asymptotic behavior of each radial channel. The network is left to learn the residual intermediate-range structure. This construction follows the physics-informed neural-network (PINN) philosophy~\citep{Karniadakis_2021,raissi2019physics}, while remaining closer to a structured variational wave function method than to a black-box PDE solver. Our main contributions are:
\vspace{-0.25em}
\begin{itemize}[leftmargin=2.5em]
    \setlength\itemsep{0em}
    \item We introduce a neural residual representation for molecular few-body wave functions that combines analytic symmetries and asymptotic structure with a learned residual;
    \item We develop a stable physics-informed initialization and grid-sampling procedure;
    \item We formulate the approach for both two- and three-body systems, and validate it for two bodies against mapped discrete variable representation (DVR) calculations~\citep{Willner2004} and for three bodies against a two-dimensional DVR calculation;
\end{itemize}
\vspace{-0.25em}
\textbf{Extensibility to larger systems:} 
In few-body physics, the particle count $(N)$ indexes a family of related $N$-body problems, whose complexity scales exponentially with $N$. While each~$N$ requires a corresponding wave function formulation, the principles underlying our representation remain unchanged. Translational invariance is enforced using Jacobi coordinates, while the angular dependence of multiple relative coordinates is represented through tensor products of spherical harmonics and standard angular-momentum coupling~\citep{Varshalovich_Book}. These ingredients allow the representation to extend systematically to larger few-body systems, although the computational complexity still grows rapidly with $N$. We formulate and test our method on two- and three-body systems. For two bodies, we benchmark it against an established high-accuracy reference solution and conduct most of our ablations and analyses of individual design choices. For three bodies, we obtain ground-state energies in percent-level agreement with an independent grid-based calculation. Field-linked three-body bound-state energies remain less established, with recent calculations by \citet{singh2026efimov} providing an initial study. Hence, the three-body comparison measures numerical agreement between independent solvers rather than error relative to the established ground truth. More broadly, our results illustrate the benefit of using analytical physical structure to reduce the learning burden of neural wave function representations.

\section{Related Work}

\textbf{Neural few-body solvers.} Neural approaches for continuous-coordinate few-body systems remain underexplored relative to their many-body counterparts. Early works used feedforward networks for few-body bound states of atoms, focusing on universal phenomena and small $^{4}\mathrm{He}_N$ clusters~\citep{Saito_2018,Yokoi_2026,freitas2023synergy}, demonstrating neural representations for a few structureless particles interacting through isotropic short-range potentials. Other works \citep{Keeble_2020,Rozalen_Sarmiento_2024,Li_2025} extended NNs to few-nucleon systems with spins, though still under simple short-range interactions. In contrast, we focus on dipolar molecular systems, where anisotropic long-range interactions couple internal degrees of freedom. Section~\ref{sec_challenges} details the numerical demands that this interaction profile imposes on the solver.

\textbf{Physics-informed PDE solvers.} Our work also connects to neural PDE solvers~\citep{raissi2019physics}, variational approaches, like Deep Ritz~\citep{yu2018deep}, and extensions to eigenvalue problems and operator learning~\citep{jin2022physics,lifourier,lu2021learning}.
However, standard neural PDE solvers rarely incorporate the physical symmetries, asymptotic behavior, or variational structure intrinsic to quantum few-body Hamiltonians, although some recent approaches explicitly encode Lie-group symmetries~\citep{Klausen_2025}.

\textbf{Symmetry-aware architectures.} Our approach aligns with symmetry-aware architectures that embed geometric priors via equivariant NNs to improve efficiency and generalization~\citep{thomas2018tensor,kondor2018generalization,batzner20223}.
Our method follows this philosophy by combining variational learning with systematic physics-informed structure tailored to first-quantized few-body Schr\"odinger problems.

\section{Background}
\label{sec_background}

\noindent\textbf{Wave function and Hamiltonian.}
The wave function $\ket{\Psi}\sim\Psi(\bm{\alpha})$, defined over generalized coordinates~$\bm{\alpha}$, fully describes the quantum state of a \mbox{physical system~\citep{Sakurai_Book}.} However, $\ket{\Psi}$ itself is not directly observable. Instead, according to the \mbox{Born rule~\citep{Born_1926},} the normalized wave function defines the \textit{probability density}: ${p(\bm{\alpha})=|\Psi(\bm{\alpha})|^{2}}$. Quantum observables are represented by Hermitian operators~$\hat{O}$ acting on~$\Psi(\bm{\alpha})$, whose expectation values are computed with respect to the probability density $p(\bm{\alpha})$.
One such operator is the Hamiltonian $\hat{\mathcal{H}}(\bm{\alpha};\bm{\eta})$, which represents the total energy of the system $\mathcal{E}$ and may depend on extrinsic parameters $\bm{\eta}$. Expectation values are computed as: $\mathcal{E}=\bra{\Psi}\hat{\mathcal{H}}\ket{\Psi}=\int \Psi^{\ast}(\bm{\alpha})\{\hat{\mathcal{H}}\Psi\}(\bm{\alpha})\,\mathrm{d}\bm{\alpha}$, where $\Psi^{\ast}(\bm{\alpha})$ \mbox{is a complex conjugate.}

\noindent\textbf{Schr\"odinger equation.} Our goal is to numerically approximate \textit{stationary states} of a physical system. These states are eigenstates of the Hamiltonian operator $\hat{\mathcal{H}}(\bm{\alpha};\bm{\eta})$, and are solutions to the time-independent Schr\"odinger equation~\citep{Sakurai_Book,Schroedinger_1926}
\begin{equation}
    \hat{\mathcal{H}}(\bm{\alpha};\bm{\eta})\Psi_{n}(\bm{\alpha})=\mathcal{E}_{n}\Psi_{n}(\bm{\alpha})\,,
    \label{eq_Schroedinger}
\end{equation}
where index $n$ labels states~$\Psi_{n}(\bm{\alpha})$ with eigenenergies $\mathcal{E}_{n}$, and $\bm{\eta}$ denotes extrinsic parameters.

\noindent\textbf{Variational method.}
\label{sec_variational_method}
The conventional numerical approach to solving eigenvalue problems, such as~Eq.~\eqref{eq_Schroedinger}, employs the variational principle~\citep{Yuan_2019}. One considers an arbitrary function~$\ket{\Psi_{\bm{w}}}$, parametrized in a complete functional basis $\ket{\psi^{w}_k}$, such that $\ket{\Psi_{\bm{w}}}=\sum_{k=1}^{\infty}w_k\ket{\psi^{w}_k}$.
The~normalized expectation value of the Hamiltonian (or Rayleigh quotient) is
%Minimization over coefficients $\bm{w}$ yields the normalized expectation value of the Hamiltonian
\begin{equation}
    \mathcal{E}_{\bm{w}}[\hat{\mathcal{H}}]\equiv\langle\hat{\mathcal{H}}\rangle=\bra{\Psi_{\bm{w}}}\hat{\mathcal{H}}\ket{\Psi_{\bm{w}}}
    /
    \bra{\Psi_{\bm{w}}}\ket{\Psi_{\bm{w}}}\,.
    \label{eq_energy_functional}
\end{equation}
The parameters $\bm{w}_0$ that minimize $\mathcal{E}_{\bm{w}}[\hat{\mathcal{H}}]$ yield the lowest-energy eigenstate of $\hat{\mathcal{H}}(\bm{\alpha};\bm{\eta})$ within the chosen parametrization. Its energy, $\mathcal{E}_{0}=\mathcal{E}_{G}=\mathcal{E}_{\bm{w}_{0}}[\hat{\mathcal{H}}]$, is called the \textit{ground-state} energy.

Higher-energy (excited) states can be obtained via the Gram--Schmidt orthogonalization~\citep{MacDonald_1933,Ding_2024}. Given $n_0$ lowest-energy wave functions $\{\ket{\Psi^{0}},...,\ket{\Psi^{n_0-1}}\}$, one can repeat the variational algorithm in the subspace orthogonal to these functions using \mbox{the trial wave function}
\begin{equation}
    \ket{\Psi_{\bm{w}}} = \ket{\Psi^{n_0}_{\bm{w}}} - \sum\nolimits_{n=0}^{n_0-1}\big(\bra{\Psi^{n}}\ket{\Psi^{n_0}_{\bm{w}}}/\bra{\Psi^{n}}\ket{\Psi^{n}}\big)\ket{\Psi^{n}}\,.
    \label{eq_orthogonalization}
\end{equation}
Eq.~\eqref{eq_orthogonalization} is not the only strategy for calculating excited states~\citep{Gonzalez_2020}. Alternatives include secular determinant or generalized Rayleigh--Ritz method~\citep{Coulson_Book,Roothaan_1951}, used, for example, by~\citet{Pfau_2024}. These methods construct overlap matrices from trial functions and solve a generalized eigenvalue problem, yielding several low-energy states simultaneously. While useful when orthogonalization becomes numerically unstable, these methods require more resources. For our purposes, a simpler method, based on Eq.~\eqref{eq_orthogonalization}, provides sufficient performance.

\noindent\textbf{Training objective.} Eqs.~\eqref{eq_energy_functional} and \eqref{eq_orthogonalization} define the training objective used throughout this work. Implementation details and pseudocode for the optimization procedure are \mbox{provided in Apps.~\ref{app_pseudo_code} and~\ref{app_technical_details}.}

\section{Problem formulation}
\label{sec_problem_formulation}

\subsection{Physical model}
\label{sec_physical_model}

\noindent\textbf{Hamiltonian of a few-body system.}
The Hamiltonian operator $\hat{\mathcal{H}}(\bm{\alpha};\bm{\eta})$ (see Sec.~\ref{sec_background}) describes an $N$-body system with coordinates $\bm{\alpha}\equiv \{\bm{R}_{i}\}_{i=1}^N$ and physical parameters $\bm{\eta}\equiv\bm{\eta}_{\text{pot}}\cup\{m_i\}_{i=1}^N$. For particles with masses $\{m_i\}_{i=1}^N$, the Hamiltonian partitions into kinetic and potential terms:
\begin{equation}
    \hat{\mathcal{H}} = \hat{\mathcal{H}}_\text{kin} + \hat{\mathcal{H}}_\text{pot}
    \,\,,\,\text{where }\quad
    \hat{\mathcal{H}}_\text{kin} = \sum\nolimits_{i=1}^{N}\mathcal{E}^{\text{kin}}_{i}\bm{\nabla}^2_{\bm{R}_i}
    \quad\text{and}\quad
    \hat{\mathcal{H}}_\text{pot} = \sum\nolimits_{i<j}^N V(\bm{\eta}_{\text{pot}},\bm{R}_{i}-\bm{R}_{j})\,.
    \label{eq_hamiltonian_few_body}
\end{equation}
Here ${V(\bm{\eta}_{\text{pot}},\bm{R}_{i}-\bm{R}_{j})}$ is the interaction between $i$-th and $j$-th particles, $\bm{\nabla}_{\bm{R}_i}$ is the gradient operator, and $\mathcal{E}^{\text{kin}}_{i}=-\hbar^2/(2m_i)$ are kinetic factors, where $\hbar$ is the reduced Planck constant. We assume that all particles have the same mass $m$, and that ${V(\bm{\eta}_{\text{pot}},\bm{R}_{i}-\bm{R}_{j})}$ depends only on the relative position vector $\bm{R}_{i}-\bm{R}_{j}$. Such an interaction implies overall \textit{translational invariance} in the system, which is characteristic of systems without inhomogeneous external fields.

For translationally invariant systems, we replace laboratory coordinates $\{\bm{R}_{i}\}_{i=1}^N$ with Jacobi relative coordinates $\{\bm{r}_{i}\}_{i=1}^{N-1}$ (defined in App.~\ref{app_Jacobi}, see \citet{Suzuki_1998} and \citet{Jacobi_Lectures}) and the center-of-mass coordinate $\bm{R}_{\mathrm{CM}}$. $\bm{R}_{\mathrm{CM}}$ describes trivial global motion, separate from the intrinsic dynamics, and can be discarded. We therefore solve only for $\bm{\Psi}(\bm{r}_1,\ldots,\bm{r}_{N-1})$, reducing the dimensionality while explicitly enforcing translational invariance on the level of the Hamiltonian.

\begin{figure}[h]
    \centering
    %\printsizes
    \vspace{-1.em}
    \begin{minipage}[c]{0.33\textwidth}
        \caption{\textbf{Interaction potential of field-linked molecules.} (a)~Potential $V(\bm{r})$ is strongly anisotropic with respect to the angle $\theta$, yielding intricate correlation dynamics. (b)~Potential $V(\bm{r})$ exhibits energy barrier for short-range $(|\bm{r}|\to0)$ interactions, whose position and strength depend on the Rabi parameter $\nu$. 
        % \JL{I suggest switching back to $\Omega$ for Rabi frequency, as it is in the appendix, to be more conventional. And the number shown in the figure and everywhere is $\Omega/2 \pi$. Or maybe an easier solution is to define $\nu=\Omega/2\pi$ in the appendix so that you don't need to modify figures. We also need to say somewhere we take $^{20}$Na$^{40}$K as an example because then the value of potential makes sense.}
        }
        \label{fig_potential}
    \end{minipage}%
    \hfill
    \begin{minipage}[c]{0.65\textwidth}
        \vspace{0.25em}
        \centering
        %\printsizes
        \includegraphics[width=\linewidth]{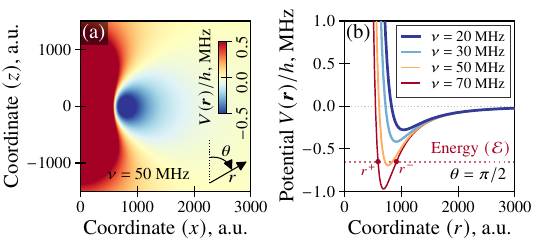}
    \end{minipage}
    \vspace{-1.em}
\end{figure}

\noindent\textbf{Field-linked molecules.}
We focus on ultracold polar molecules. When these molecules approach each other at short range, they undergo inelastic or reactive processes \citep{Ospelkaus:2010,Liu:2020,Gregory:2020,Bause:2023}, which leads to clustering and loss of correlations.
%Mayle:2013,Christianen:2019,
%We focus on anisotropic pairwise interactions arising from ultracold dipolar molecules dressed by a microwave electric field~\cite{Chen_2023}. Such molecules are described by the following interaction 
%As anisotropic long-range dipole–dipole interactions are attractive for certain orientations,  
To overcome this challenge, an external (microwave) field can be coupled to molecules to induce a potential barrier that strongly suppresses short-range losses, a mechanism known as \textit{microwave shielding} \citep{Buchler:2007,Micheli:2007,Cooper2009, Karman2018,Anderegg:2021,Deng_2023}. %Lassabliere2018,Huang:2012,
The potential barrier manifests itself in a rapid increase of the interaction energy ${V(\bm{\eta}_{\text{pot}},\bm{r})}$, as $|\bm{r}|\to 0$ (Fig. \ref{fig_potential}(b)). The shielding potential between microwave-dressed molecules is well approximated by the following expression~\citep{Deng_2023}:
\begin{equation}
    V(\bm{\eta}_{\text{pot}},\bm{r})=C_6(\bm{\eta}_{\text{pot}}) [1-\cos^4\theta]\,r^{-6} + C_3(\bm{\eta}_{\text{pot}})[3\cos^2 \theta-1]\,r^{-3}\,,
    \label{eq_potential}
\end{equation}
where $\bm{r}=\{r,\theta,\phi\}$ denote spherical coordinates, and interaction strength coefficients $C_3$ and $C_6$ depend on molecular properties and can be tuned via field parameters~$\bm{\eta}_{\text{pot}}=\{\nu,...\}$. Their explicit form is provided in App.~\ref{app_field_linked}. The main extrinsic physical parameter is the Rabi coupling~$\nu$ that can be adjusted to shift the barrier to shorter range (Fig.~\ref{fig_potential}(b)), deepening the potential well. When depth of the well is sufficient to fit a matter-wave mode, the potential supports molecular binding. Such a molecular bond is stable only in the presence of the external electric field and is therefore referred to as a \textit{field-linked} bound state. Recent observation of field-linked dimer states in dipolar molecules~\citep{Chen:2024} demonstrates that these bonds can be reversibly created and destroyed by tuning the external field, enabling controllable assembly of larger multi-molecular bound states.

\subsection{Challenges for numerical approaches}
\label{sec_challenges}

The physical model introduced in Sec.~\ref{sec_physical_model} poses several challenges for eigenvalue solvers. It requires simultaneously resolving continuous spatial coordinates and intricate interaction-induced correlations.

Quantum wave function representations are often tailored to two limiting regimes: many-body problems with established ansätze or analytically tractable few-body systems.
%Quantum wave function representations are often tailored to either many-body systems with established ans\"atze or simple few-body systems admitting closed-form solutions.
In many-body physics (electronic or spin states), single-particle states can often be represented in truncated bases, enabling physically motivated multi-particle trial states, like Slater determinants~\citep{Pfau_2020,Hermann_2020}. These approaches efficiently capture correlations, but become increasingly difficult to construct as the complexity of single-particle states grows. At the other extreme are simple few-body systems, such as harmonic confinement or isotropic interactions, where closed-form solutions are often available. These methods typically focus on the continuous nature of particle coordinates, but do not embed correlations into the trial state. Our problem lies between these regimes: although involving only a few particles, the interactions are long-range and anisotropic~(see Fig.~\ref{fig_potential}(a)), leading to intrinsically coupled continuous degrees of freedom. As a result, the wave function does not admit a simple factorized form or compact low-dimensional parametrization, and must instead be approximated as the solution of a high-dimensional PDE eigenvalue \mbox{problem on a continuous manifold.}

The interaction potential~\eqref{eq_potential} is particularly difficult for numerical representation. First, the external microwave field, which induces the shielding, defines a preferred axis; and the interaction depends on the orientation of the intermolecular separation relative to that axis. This anisotropy couples different partial waves, so the solver must resolve several radial-channel wave functions simultaneously. Second, short-range shielding and the long-range dipolar attraction set distinct radial scales. The solver must resolve rapid variation near the repulsive core, while retaining a radial domain large enough to capture the extended tails of weakly bound states. These demands change with the Rabi frequency $\nu$~(see Fig.~\ref{fig_potential}(b)): deeper wells produce more localized states, while weakly bound excited states require the solver to resolve both radial nodes and an extended outer tail.

%Interaction potential~\eqref{eq_potential} is particularly difficult for numerical representation. On the one hand, the Hamiltonian~\eqref{eq_hamiltonian_few_body} is invariant under global translations and rotations, implying that a physically meaningful solution to Eq.~\eqref{eq_Schroedinger} needs to be symmetric with respect to the Euclidean group~E$(3)$. On the other hand, anisotropic potential $V(\bm{\eta}_{\text{pot}},\bm{r})$ results in single particle states that violate the rotational symmetry of SO$(3)$, a subgroup of E$(3)$. Moreover, the competing length scales, arising from $r^{-3}$ and $r^{-6}$ terms in Eq.~\eqref{eq_potential}, induce both sharp and slowly varying radial structure in the wave function, particularly for excited states with an additional nodal structure.

These challenges call for a flexible representation of a wave function, which is capable of capturing the rich structure of solutions to Eq.~\eqref{eq_Schroedinger} within a unified framework. Such a representation should avoid restrictive functional bases, as it has to adapt to the structural diversity of the eigenfunctions. The variational formulation introduced in Sec.~\ref{sec_background} applies to any parametric model. In this work, we adopt a neural network parametrization of the wave function, augmented with physics-informed inductive biases to capture the underlying structure of the problem.

%\section{Representation of the Wave Function}
\section{Physics-Informed Neural Variational Method}

% In this section, we introduce a neural wave function representation that embeds the geometric and physical structure of the problem, described in Sec.~\ref{sec_physical_model}. The \textit{central idea} is to encode symmetry and anisotropic behavior, into the representation itself, thereby removing the associated complexity from the learning problem. As a result, the task reduces to approximating a low-dimensional residual within a unified variational formulation, which substantially reduces computational demand from the neural network. We first describe the symmetry-inspired angular decomposition (Sec.~\ref{sec_spherical}), followed by the analysis of asymptotic behavior (Sec.~\ref{sec_asymptotic}), and sampling strategy (Sec.~\ref{sec_sampling}).

We introduce a neural wave function that embeds the geometric and physical structure of the problem, described in Sec.~\ref{sec_physical_model}. 
The \textit{key idea} is to encode relevant symmetries into the representation, reducing the problem to learning a low-dimensional residual within a unified variational formulation.

%We first describe the angular decomposition~(Sec.~\ref{sec_spherical}), followed by the asymptotic analysis (Sec.~\ref{sec_asymptotic}), and sampling strategy (Sec.~\ref{sec_sampling}).

%The final representation of the wave function is
%$
%{\Psi}(\bm{r}) = \sum_{l,m} [\psi_{l,m}^{\text{NN}}(r)\psi_{\text{WKB}}(r)/r]\,Y_{l,m}(\theta,\phi),
%$
%where $\psi_{\text{WKB}}(r)$ is a shared asymptotic envelope and $\psi_{l,m}^{\text{NN}}(r)$ are learnable neural residuals.

% This section contains the core information about our method. In particular, it delineates the representation that we use to reduce the dimensionality of the problem. 
%n this section, we introduce a representation that takes account of the geometric and physical structure of the problem directly at the level of the wave function. The central idea is to encode known properties, such as symmetry and asymptotic behavior, into the representation itself, thereby removing much of the associated complexity from the learning problem. As a result, the task reduces to approximating a low-dimensional residual radial component within a unified variational formulation, significantly reducing the burden on the neural network.

\subsection{Spherical decomposition}
\label{sec_spherical}

The action of the Hamiltonian operator~\eqref{eq_hamiltonian_few_body} on the wave function has an internal \mbox{symmetry structure.} While this structure can be learned by the network~(see Sec.~\ref{sec_experiments_ablations} and App.~\ref{app_2d_training}), doing so is highly inefficient when this angular structure is analytically known and can be embedded into the representation~\citep{Klausen_2025}. We therefore encode the SO$(3)$ angular-momentum coupling rules into the representation for $\bm{\Psi}(\bm{r}_1,...,\bm{r}_{N-1})$, treating molecular rotations purely analytically. 
Full derivations for two- and three-molecule systems are presented respectively in Apps.~\ref{app_two_body} and~\ref{app_three_body}.
In the 2-body case, we expand the wave function in spherical \mbox{harmonics~\citep{Varshalovich_Book}:}
$
    {\Psi}(\bm{r}) = \sum_{l,m} [\psi_{l,m}(r)/r]\,Y_{l,m}(\theta,\phi),
    \label{eq_spherical_expansion}
$
where $\bm{r}=\{r,\theta,\phi\}$ are spherical coordinates and $Y_{l,m}(\theta,\phi)$ are spherical harmonics. 
Matrix elements of the Hamiltonian become: $\mathcal{H}_{l,m;l',m'}(r)=\bra{Y_{l',m'}}\hat{\mathcal{H}}\ket{Y_{l,m}}$, reducing the Schr\"odinger equation~\eqref{eq_Schroedinger} to a leaner system of coupled radial equations
\begin{equation}
    \sum\nolimits_{l',m'} \mathcal{H}_{l,m;l',m'}(r) \, \psi_{l',m'}(r) = \mathcal{E}\,\psi_{l,m}(r)\,.
    \label{eq_Schroedinger_radial}
\end{equation}

Spherical harmonics~$Y_{l,m}(\theta,\phi)$ offer several advantages as an angular basis.
First, they form irreducible representations of the SO$(3)$ group and provide a natural orthonormal basis for angular functions. Unlike problem-specific polynomial bases, they arise directly from rotational symmetry and therefore remain applicable under changes to interactions or boundary conditions.
Second, they analytically absorb singularities of spherical coordinates and diagonalize the kinetic energy operator~\eqref{eq_hamiltonian_few_body} in the angular basis, yielding numerically stable differential operators and reducing computational cost.
Third, the decomposition naturally extends to higher-body systems through tensor products of spherical harmonics and angular-momentum coupling rules, enabling scalable representations of multi-coordinate wave functions.
Finally, the expansion admits a physically justified truncation at finite $l_{\max}$, since higher-$l$ channels are suppressed by centrifugal barriers and therefore contribute negligibly to the dynamics.
Further discussion is provided in App.~\ref{app_spherical_motivation}.

\subsection{WKB asymptotic as an inductive bias}
\label{sec_asymptotic}
%A common approach is to incorporate multiplicative envelope functions, such as Jastrow factors, which enforce decay and short-range structure. However, such constructions typically require problem-specific design and do not directly reflect the underlying Hamiltonian.

% After incorporating the symmetry structure of the Hamiltonian into the wave function representation, to further increase the efficiency of the method, we encode the asymptotic behavior of radial functions~$\psi_{l,m}(r)$ into the representation as well. The expression for asymptotics can be obtained by performing multiple-scale analysis~\cite{} of the differential equation~\eqref{eq_Schroedinger_radial}, which results in the Wentzel-Kramers-Brillouin (WKB) approximation to $\psi_{l,m}(r)$. 

After incorporating the symmetry structure of the Hamiltonian into the wave function representation, we further improve efficiency by encoding the asymptotic behavior of radial functions~$\psi_{l,m}(r)$. The asymptotic form is obtained through multiple-scale analysis~\citep{Bender_Book} of Eq.~\eqref{eq_Schroedinger_radial}, yielding the Wentzel--Kramers--Brillouin (WKB) approximation~\eqref{eq_WKB}. 
Specifically, in the ``forbidden'' region, where $V(r)>\mathcal{E}$ (see Fig.~\ref{fig_potential}(b)), the wave function can be approximated by
\begin{equation}
    \psi_{\text{WKB}}(r,\mathcal{E})\sim \exp(\pm\int_{r_{\pm}}^{r}\kappa_{\mathcal{E}}(x)\mathrm{d}x)\,
    \text{, where}\quad
    \kappa_{\mathcal{E}}(r)=\sqrt{(V(r)-\mathcal{E})/|\mathcal{E}_{\text{kin}}|}\,,
    \label{eq_WKB}
\end{equation}
where $\mathcal{E}$ is the energy of the system, $V(r)\equiv V(r,\theta=\tfrac{\pi}{2},\phi=0)$ (see Eq.~\eqref{eq_potential}) and the sign is ``$+$'' for $r\ll r_{+}$ and ``$-$'' for $r\gg r_{-}$, where $r_{\pm}$ are the turning points, defined as solutions to $\kappa_{\mathcal{E}}(r)=0$ (see Fig.~\ref{fig_potential}(b)). Approximation Eq.~\eqref{eq_WKB} is rather inaccurate around and between the turning points~$r_{\pm}$, however it provides the correct asymptotics in $r\to0$ and $r\to\infty$ regimes.

\begin{wrapfigure}{r}{0.48\textwidth}
    \vspace{-1.5em}
    \centering
    %\printsizes
    \includegraphics[width=0.46\textwidth]{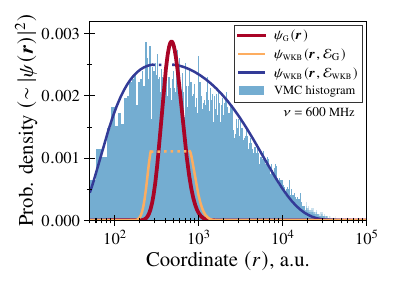}
    \vspace{-1.5em}
    \caption{
    \textbf{WKB-inspired asymptotic envelope used for sampling.}
    The approximation $\psi_{\mathrm{WKB}}(r,\mathcal{E}_{\text{G}})$ captures the asymptotic behavior of the wave function $\psi_{\mathrm{G}}(r)$. To improve convergence, we use a broader reference envelope $\psi_{\mathrm{WKB}}(r,\mathcal{E}_{\text{WKB}})$ as a prior and as a fixed sampling distribution for the VMC. The histogram shows VMC samples drawn from this envelope.
    }
    %\caption{\textbf{Asymptotic envelope inspired by the WKB approximation and and used as a sampling distribution.} Approximation $\psi_{\text{WKB}}(r,\mathcal{E}_{\text{G}})$ accurately models asymptotic behavior of the wave function $\psi_{\text{G}}(r)$. To ensure convergence, a broader reference envelope $\psi_{\text{WKB}}(r,\mathcal{E}_{\text{ref}})$ can be used as a prior. The same envelope can be used to generate VMC samples.
    %Histogram of VMC samples from the WKB-inspired envelope, with reference $\psi_{0,0}(r)$ and WKB approximation $\psi_{\mathrm{WKB}}(r)$. The fixed distribution captures large-scale (asymptotic) structure.
    %}
    \label{fig_WKB}
    \vspace{-3.5em}
\end{wrapfigure}

Since the exact system energy $\mathcal{E}$ is unknown before the training, we define the reference envelope using the proxy energy $\mathcal{E}_{\text{WKB}}=\min_r V(r)/2$. This proxy does not approximate the ground-state energy, in fact, for $\nu=50\,\text{MHz}$: $\mathcal{E}_{\text{WKB}}\approx10\,\mathcal{E}_{\text{G}}$. Instead, it tracks changes to interaction parameters, such as $\nu$. Altogether, we use WKB-inspired asymptotic envelope $\psi_{\text{WKB}}(r,\mathcal{E}_{\text{WKB}})$ instead of employing the approximation itself.

Figure~\ref{fig_WKB} provides comparison between typical behavior of target ground-state wave function $\psi_{\text{G}}(r)$ and the corresponding WKB approximations~\eqref{eq_WKB}. A literal WKB envelope can be overly restrictive, limiting the neural network's freedom to explore nearby parameters. Hence, we intentionally weaken the asymptotic prior by broadening the envelope via $\kappa_{\mathcal{E}}(r)\to\beta\kappa_{\mathcal{E}}(r)$, where $\beta\in(0,1)$. The details can be found in  App.~\ref{app_technical_details}.

\subsection{Neural residual}

%Overall, the decomposition treats angular dependence analytically through spherical harmonics, leaving only radial functions $\psi_{l,m}^{\text{NN}}(r)$ to be learned by a simple NN - multi-layer perceptron (MLP). The same network weights are \textit{shared} across all angular channels, enabling efficient inter-channel coupling in a lower-dimensional radial domain with a simple architecture.

%To sum up, in the two-body case, the wave function is represented by
%\begin{equation}
%    {\Psi}(\bm{r}) = \sum_{l,m} [\psi_{l,m}^{\text{NN}}(r)\psi_{\text{WKB}}(r)/r]\,Y_{l,m}(\theta,\phi),
%\end{equation}
%where $Y_{l,m}(\theta,\phi)$ are angular channels, discussed in Sec.~\ref{sec_spherical}, $\psi_{\text{WKB}}(r)$ is a shared asymptotic envelope, introduced in Sec.~\ref{sec_asymptotic}, and $\psi_{l,m}^{\text{NN}}(r)$ are learnable neural residuals.

Combining the angular decomposition of Sec.~\ref{sec_spherical} with the asymptotic envelope of Sec.~\ref{sec_asymptotic}, the neural network only needs to represent the remaining radial structure. We parametrize these residual radial functions with a \textit{shared} multi-layer perceptron (MLP), which maps the radial coordinate $r$ jointly to all angular channels $\psi_{l,m}^{\text{NN}}(r)$. Sharing the network parameters across channels allows correlations between different angular components to be captured within a common low-dimensional representation. For the two-body system, the resulting wave function representation is
\begin{equation}
    {\Psi}(\bm{r}) =
    \sum_{l,m}
    [\psi_{l,m}^{\text{NN}}(r)\,\psi_{\text{WKB}}(r, \mathcal{E}_{\text{WKB}})/r]\,
    Y_{l,m}(\theta,\phi)\,.
    \label{eq_two_body_representaiton}
\end{equation}
The network parameters are optimized variationally using the objective of Sec.~\ref{sec_background}. For the ground state, we minimize the normalized expectation value of the Hamiltonian~\eqref{eq_energy_functional}, while excited states are obtained by applying the same optimization to the orthogonalized trial wave function~\eqref{eq_orthogonalization}. At each optimization step, the neural residual $\psi_{l,m}^{\text{NN}}(r)$ is evaluated on a minibatch of radial coordinates, combined with the fixed analytical components to construct the full wave function~\eqref{eq_two_body_representaiton}, and used to evaluate the variational energy. Gradients are then backpropagated through the neural residual. Thus, only the residual functions are learned, while the angular basis and asymptotic envelope remain fixed. The complete procedure and implementation details are provided in Apps.~\ref{app_pseudo_code} and~\ref{app_technical_details}.

\subsection{Sampling and Grid Construction}
\label{sec_sampling}

Accurate evaluation of the variational objective depends critically on the sampling strategy: regions that contribute substantially to the expectation value must be sufficiently resolved to avoid biased energy estimates. Ideally, the sampling grid should therefore follow the underlying wave function distribution. A straightforward approach uses a fixed discretization of coordinate space. However, the wave function exhibits strong multi-scale structure~\citep{Bender_Book}, with sharp localization in some regions and decay over many orders of magnitude elsewhere. Uniform grids are therefore inefficient: they either fail to resolve narrow features or allocate excessive resolution to regions with negligible probability mass (see App.~\ref{app_exp_sampling}). Fixed coordinate transformations improve resolution across scales, but remain problem-specific and nonadaptive.

This motivates sampling from a target probability distribution. In variational Monte Carlo (VMC), samples are drawn from the model-induced density $\rho_{\bm{\omega}}(\bm{r})\sim|\Psi_{\bm{\omega}}(\bm{r})|^{2}$ using Markov chain Monte Carlo (MCMC) methods~\citep{Brooks_Book,Chib_1995,Robert_Book}. Because this distribution evolves during optimization, the Markov chain must be repeatedly re-equilibrated, increasing computational cost and destabilizing training. Moreover, the learned distribution may transiently become nearly uniform (Sec.~\ref{sec_exp_dynamics}), making it uninformative for sampling. We therefore decouple grid construction from the evolving model by initializing a fixed MCMC sampling grid before training. Specifically, we sample points from a target density proportional to $|\psi_{\text{WKB}}(r,\mathcal{E}_{\text{WKB}})|^2$, introduced in Sec.~\ref{sec_asymptotic}, and draw minibatches from this fixed grid during optimization. 
As shown in Fig.~\ref{fig_WKB}, the resulting distribution captures the large-scale structure of the problem while remaining independent of the evolving model. Altogether, the fixed sampling ensures stable and accurate training, as discussed in App.~\ref{app_exp_sampling}. Importantly, the samples only define a non-uniform grid. After sorting the grid points, the integrals in Eqs.~\eqref{eq_energy_functional} and \eqref{eq_orthogonalization} are evaluated using the trapezoidal rule: $\int f\,\mathrm{d}r \approx \sum_i w_i f(r_i)$ with weights $w_i = (r_{i+1}-r_{i-1})/2\propto 1/|\psi_{\text{WKB}}(r_i,\mathcal{E}_{\text{WKB}})|^2$ on average. Implementation details are provided in App.~\ref{app_technical_details}.

\section{Experiments}

\begin{wrapfigure}{r}{0.57\textwidth}
    \vspace{-4.em}
    \centering
    \includegraphics[width=0.55\textwidth]{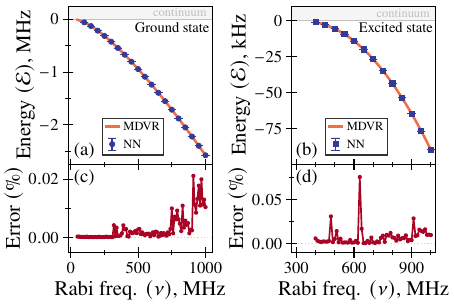}
    \vspace{-0.5em}
    \caption{
    \textbf{Comparison with MDVR.}
    Predictions from our approach vs. MDVR reference for ground and excited states across  $\nu$. Top: predicted and reference energies. Bottom: relative absolute errors $\Delta_{\mathcal{E}}$.
    The model accurately reproduces the energy spectrum across a broad range of interaction regimes.
    }
    \label{fig:main_results}
    \vspace{-2.em}
\end{wrapfigure}

We evaluate our method on two-body field-linked molecular systems, where high-accuracy reference solutions are available. Our experiments address four questions: (i) can the method reproduce MDVR reference solutions, (ii) which components of the physics-informed design are essential, (iii) do the core numerical building blocks for higher-body extensions function reliably, (iv) how does optimization behave across interaction regimes? All experiments use a shared model configuration \textit{without fine-tuning} and exhibit stable optimization behavior. Extended results are provided in App.~\ref{app_extended_experiments}, with implementation details and pseudocode in Apps.~\ref{app_technical_details} and~\ref{app_pseudo_code}, respectively.

\subsection{Comparison with MDVR}
\label{exp_main}
We compare our approach against MDVR (App.~\ref{app_dvr}), a high-accuracy grid-based variational solver widely used for molecular bound-state calculations (which we independently validate against a Weinberg-expansion benchmark in App.~\ref{app_exp_weinberg}). We evaluate ground and excited states across Rabi frequencies $\nu \in [50,1000]\,\text{MHz}$ ($\Delta\nu = 10\,\text{MHz}$). 
Sweeping $\nu$ alters the depth and position of the field-linked potential well, probing physically distinct optimization regimes: low $\nu$ produces shallow wells with spatially extended states and weak gradients, whereas high $\nu$ induces strong confinement and tightly localized wave functions. Excited states appear only for $\nu \gtrsim 400\,\text{MHz}$, as lower frequencies lack a sufficiently deep potential well to support bound excited states.

The primary metric that we use to quantify the agreement between the neural and reference energies is the relative discrepancy in percents:
\begin{equation}
\Delta_{\mathcal{E}}[\%]=100 \cdot |\mathcal{E}_{\text{neural}}-\mathcal{E}_{\text{reference}}| / |\mathcal{E}_{\text{reference}}|.
\label{eq_relative_error}
\end{equation}

As shown in Fig.~\ref{fig:main_results}, our model closely matches the MDVR reference energies across all physical regimes, accurately reproducing both the ordering and energy spacing of bound states. The mean absolute relative error is $\Delta_{\mathcal{E}}=0.003\%$ for the ground state and $\Delta_{\mathcal{E}}=0.008\%$ for excited states. The solver accurately reproduces the excited-state energy even near its emergence threshold ($\nu \approx 400\,\text{MHz}$), where the state is only weakly bound. Multi-seed experiments demonstrate low variance and high stability across initialization seeds (App.~\ref{app_exp_robustness}). Although MDVR remains computationally faster in this low-dimensional setting, our goal is not to outperform specialized grid solvers, but to demonstrate that our representation robustly captures distinct physical regimes while extending to higher-dimensional few-body systems.

%\begin{figure}[t]
%    \centering
    %\printsizes
%    \includegraphics[width=0.78\linewidth]{figures/wkb_vs_no_wkb_energy_by_nu_row_ground_state.pdf}
%    \caption{\textbf{Effect of the WKB-induced asymptotic bias.}
%Training trajectories with and without the WKB-inspired envelope for the ground state across different $\nu$. Embedding the correct asymptotic decay directly into the wave function representation accelerates convergence, with the largest improvement at low $\nu$ where the wave function is spatially extended.}
%    \vspace{-1.5em}
%    \label{fig_experiment_WKB}
%\end{figure}

% add number of step to convergence? mae? std?
\subsection{Ablations of Physics-Informed Components}
\label{sec_experiments_ablations}

%\begin{wrapfigure}{l}{0.48\textwidth}
%    \vspace{-2em}
%    \centering
    %\printsizes
%    \includegraphics[width=0.46\textwidth]{experiment_asymptotic.pdf}
%    \vspace{-1.em}
    %\caption{\textbf{Effect of the WKB-induced asymptotic bias.}
%Training trajectories with and without the WKB-inspired envelope for the ground state across different $\nu$. Embedding the correct asymptotic decay directly into the wave function representation accelerates convergence, with the largest improvement at low $\nu$ where the wave function is spatially extended.}
%    \caption{\textbf{Effect of the WKB-inspired asymptotics.}
%Ground-state training trajectories with and without the WKB-inspired envelope across different $\nu$. Encoding the correct asymptotic decay directly into the wave function representation accelerates convergence, especially at low~$\nu$, where the wave function is spatially extended.}
%    \label{fig_experiment_WKB}
%    \vspace{-1.em}
%\end{wrapfigure}

We isolate the contribution of the main design choices by modifying individual components while keeping the variational objective and optimizer fixed. Specifically, we compare the full model against (i) a variant without the WKB-inspired asymptotic envelope, (ii) a direct matrix parametrization without shared structure across channels, and (iii) a 2D representation without analytical spherical harmonics.
Additional experiments are provided in App.~\ref{app_extended_ablation}.

\begin{figure}[h]
    \centering
    %\printsizes
    \vspace{-0.5em}
    \includegraphics[width=0.88\linewidth]{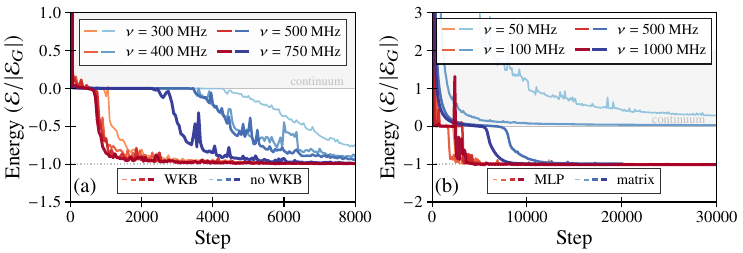}
    \vspace{-1em}
    \caption{
        \textbf{Effects of the WKB-inspired asymptotics and neural parametrization.}
        (a)~Ground-state training trajectories with and without the WKB-inspired envelope across different $\nu$. Encoding the asymptotic decay directly into the wave function representation accelerates convergence. (b)~Training dynamics of a direct matrix parametrization versus the shared MLP parametrization across~$\nu$. The matrix baseline exhibits slower convergence and instability, particularly at low  $\nu$, while the MLP model converges reliably and reaches lower energies. In both cases, the effect is more pronounced at low~$\nu$, where the wave function is spatially extended.
    }
    \label{fig_WKB_and_model}
    \vspace{-0.5em}
\end{figure}

\textbf{Effect of the WKB-induced inductive bias.}
% We therefore use a softened WKB envelope, obtained by rescaling the decay constant $\kappa$ by a factor of $100$. This retains the intended asymptotic inductive bias while preserving gradients of sufficient magnitude for stable learning.
The WKB envelope has the largest effect at low $\nu$, where bound states are weakly confined and extend over a broad spatial region. In this regime, the envelope acts as a physics-informed inductive bias, allowing the network to focus on the intermediate-range structure instead of learning long-range behavior from optimization alone. Incorporating this prior directly into the representation substantially accelerates convergence (Fig.~\ref{fig_WKB_and_model}(a)). As  $\nu$ increases and the ground state becomes more localized, the difference between the two variants decreases, since optimization becomes dominated by the local structure of the potential well.

%\begin{wrapfigure}{r}{0.48\textwidth}
%    \vspace{-2em}
%    \centering
    %\printsizes
%    \includegraphics[width=0.46\textwidth]{experiment_model.pdf}
%    \vspace{-1.em}
    %\caption{\textbf{Effect of the WKB-induced asymptotic bias.}
%Training trajectories with and without the WKB-inspired envelope for the ground state across different $\nu$. Embedding the correct asymptotic decay directly into the wave function representation accelerates convergence, with the largest improvement at low $\nu$ where the wave function is spatially extended.}
 %   \caption{
 %   \textbf{Effect of neural parametrization.}
%Training dynamics of a direct matrix parametrization (red) versus the shared MLP parametrization (blue) across  $\nu$. The matrix baseline exhibits slower convergence and instability, particularly at low  $\nu$, while the MLP model converges reliably and reaches lower energies. The gap narrows as  $\nu$ increases and the states become more localized.
%    }
%    \label{fig_vector_vs_mlp}
%    \vspace{-1.em}
%\end{wrapfigure}

\textbf{Effect of neural parametrization.}
 We consider a direct matrix parametrization as a baseline, using $100$k parameters compared to $84$k in the proposed neural representation. Here, spherical channel coefficients are represented explicitly as an $N_r \times N_\ell$ matrix, where each entry is treated as an independent learnable parameter without shared structure across radial or angular dimensions. This corresponds to independently discretizing the radial dependence of each angular channel. To isolate the effect of the parametrization, both models are multiplied by the same WKB-inspired envelope.

Despite having more parameters, the matrix parametrization performs substantially worse than the shared MLP, particularly at low $\nu$, where it often fails to converge within the optimization budget (Fig.~\ref{fig_WKB_and_model}(b)). This reflects the difficulty of learning spatially extended states with independent parameters and weak distributed gradients. As $\nu$ increases and states become more localized, the performance gap narrows. Increasing the parameter count from $100$k to $200$k does not recover convergence in the low-frequency regime, indicating that the gap is not explained by capacity alone. These results highlight the importance of shared parametrization across coupled angular channels.

\textbf{Effect of analytical symmetries.}
\label{exp_2d}
Without the analytical angular decomposition described in Sec.~\ref{sec_spherical}, the network must learn angular structure directly from a full 2D representation~(see Fig.~\ref{fig_2D_wavefunction}(a)) in reduced cylindrical coordinates $\bm{r}=\{r_\perp,z\}$. To evaluate this representation, we construct a 2D numerical solver that samples coordinate points randomly, integrates via Delaunay triangulation, evaluates kinetic energy operators on local interpolations to maintain stability, and extends the WKB-inspired prior to a product of 2D coordinate envelopes~(see App.~\ref{app_2d_explanations}). As shown in Fig.~\ref{fig_2D_wavefunction}(b), this unconstrained 2D network eventually converges to the correct bound-state energy. Yet, it suffers from slower convergence ($\sim \times 100$ longer optimization time), requires $\times 15$ larger batch sizes, and exhibits reduced training stability under identical hardware configurations. This confirms that analytically factoring out rotational symmetry provides the efficiency and stability gains in our framework.

\subsection{Optimization Dynamics}
\label{sec_exp_dynamics}

\begin{figure}[h]
    \centering
    %\printsizes
    \vspace{-1.em}
    \begin{minipage}[c]{0.34\textwidth}
        \caption{
\textbf{Optimization dynamics across interaction regimes.}
 (a)~Evolution of the ground-state probability density during training, compared with MDVR. A reduced learning rate is used to resolve the transient localization dynamics. (b)~Normalized energy trajectories for representative $\nu$. Low $\nu$ shows a plateau before localization, and high $\nu$ converges rapidly due to stronger confinement.}
        \label{fig_training_dynamics}
    \end{minipage}%
    \hfill
    \begin{minipage}[c]{0.65\textwidth}
        \vspace{0.25em}
        \centering
        %\printsizes
    \includegraphics[width=0.96\linewidth]{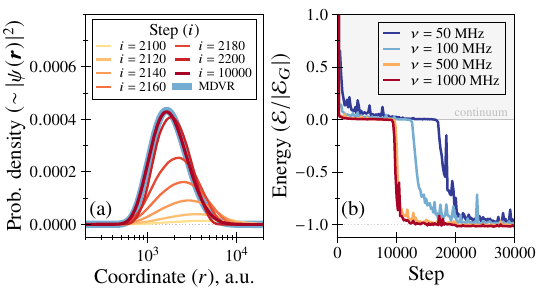}
    \end{minipage}
\end{figure}

To better understand optimization, Fig.~\ref{fig_training_dynamics} shows the evolution of energies and wave functions during training for representative low- and high-$\nu$ ground-state regimes. For small $\nu$, optimization exhibits a two-stage behavior: the model initially remains in a diffuse, near-zero-amplitude state (continuum) with weak gradients, followed by gradual accumulation of probability mass near the attractive region and rapid convergence to the bound state. Correspondingly, the wave function contracts progressively from a broad delocalized profile. In contrast, large $\nu$ induces strong gradients from the outset, leading to early localization and substantially faster convergence. In this regime, the wave function rapidly forms a localized bound state, with subsequent training only refining its structure. Overall, optimization difficulty is governed primarily by the strength of the confining potential: shallow potentials lead to slow discovery of bound states, while deep potentials enable rapid convergence.

\section{Three-Body Systems}
\label{sec_three_body}

In order to support our claim about the extension of our method to higher-body problems, we also perform calculation for a three-body $(N=3)$ system with field-linked interactions~\eqref{eq_potential}. For a three-molecule system, removing the center-of-mass motion leaves the 6D coordinate space spanned by two Jacobi vectors $\{\bm{r}_{1},\bm{r}_{2}\}$. Following the same expansion, introduced in Sec.~\ref{sec_spherical}, the 3-body wave function factorizes into dual-angular channels and coupled 2D radial residuals (see App.~\ref{app_three_body}):
\begin{equation}
    \Psi_{\text{3-body}}(\bm{r}_{1},\bm{r}_{2}) = \tfrac{1}{r_{1}r_{2}} \sum\nolimits_{l,m;L,M} \psi_{l,m;L,M}^{\text{2D}}(r_{1},r_{2}) \, \big[Y_{l,m}(\theta_{1},\phi_{1}) \otimes Y_{L,M}(\theta_{2},\phi_{2})\big]\,,
    \label{eq_3B_wavefunction}
\end{equation}
where $\bm{r}_{1}=\{r_{1},\theta_{1},\phi_{1}\}$ and $\bm{r}_{2}=\{r_{2},\theta_{2},\phi_{2}\}$ are the spherical coordinates.

\begin{figure}[h]
    \centering
    %\printsizes
    \vspace{-1em}
    \begin{minipage}[c]{0.37\textwidth}
        \caption{
            \textbf{Validation of the 2D representation.}
            (a) Ground-state probability density learned in reduced cylindrical coordinates~$\{r_{\perp},z\}$, with the sampling mesh and WKB-inspired envelope. (b) Predicted ground-state energies across $\nu$ compared with MDVR. The agreement validates the 2D radial residual required for the three-body extension.}
        \label{fig_2D_wavefunction}
    \end{minipage}%
    \hfill
    \begin{minipage}[c]{0.62\textwidth}
        \vspace{0.25em}
        \centering
        %\printsizes
    \includegraphics[width=\linewidth]{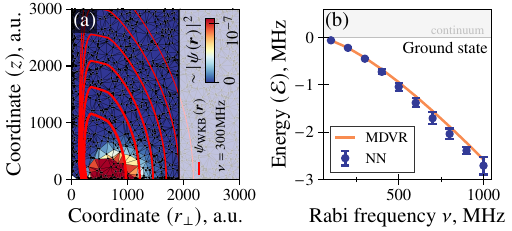}
    \end{minipage}
    %\vspace{-0.5em}
\end{figure}

In order to systematically evaluate the functionality of the representation~\eqref{eq_3B_wavefunction}, we first test the 2D radial residuals $\psi^{\text{2D}}(r_{1},r_{2})$ on the two-body problem, where there are no angular channels and an established reference MDVR benchmark is available. In particular, we consider the two molecule system in cylindrical coordinates $\{r_{\perp},z\}$, as discussed in App.~\ref{app_two_body_cylindrical}, and the network learns the ground-state wave function jointly in both coordinates. Fig.~\ref{fig_2D_wavefunction} indicates that the resulting energies agree with the MDVR across the tested Rabi frequencies $\nu$. This calculation validates the functionality of a 2D radial wave function representation.

The Hamiltonian of the three-molecule system includes three pair interactions (see App.~\ref{app_three_body}) with the same field-linked interaction profile~\eqref{eq_potential}: one acts along $\bm{r}_{1}$, while the other two act along $\bm{r}_{2}\pm\bm{r}_{1}/2$. The latter pair distances depend on both Jacobi radii $\{\bm{r}_{1},\bm{r}_{2}\}$ and their orientations, coupling the corresponding angular channels, while simultaneously introducing correlations between $\bm{r}_{1}$ and $\bm{r}_{2}$.

\begin{wrapfigure}{t}{0.42\textwidth}
    \vspace{-1.5em}
    \centering
    \includegraphics[width=0.4\textwidth]{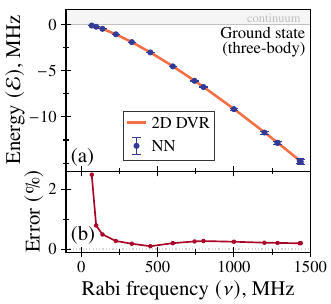}
    \vspace{-0.5em}
    \caption{
    \textbf{Three-body solver.}
    Predictions from our neural approach vs. 2D DVR reference for ground state energies across~$\nu$. (a)~Predicted and reference energies. (b)~Relative absolute errors $\Delta_{\mathcal{E}}$.
    }
    \label{fig_three_body}
    \vspace{-1.em}
\end{wrapfigure}

To perform the optimization, we minimize the expectation value of the Hamiltonian~\eqref{eq_energy_functional} on a non-uniform grid in $\{\bm{r}_{1},\bm{r}_{2}\}$ using Delaunay triangulation~\citep{Delaunay1934}. The factor $1/(r_{1}r_{2})$ in the expansion absorbs the radial Jacobian, so the energy and normalization integrals use the same triangle-area weights with measure $\mathrm{d}r_{1}\,\mathrm{d}r_{2}$. We retain angular channels up to $\{l_{\max},L_{\max}\}=\{10,8\}$. Its sufficiency is established by the ablation in App.~\ref{app_lmax}, where the cutoff increase from $\{8,8\}$ to $\{10,8\}$ enlarges the basis from 145 to 190 channels and changes the ground-state energy by at most $0.25\%$ at the frequencies tested.

Figure~\ref{fig_three_body} compares the three-body ground-state energies with an independent 2D DVR calculation across Rabi frequencies~$\nu$. Both solvers use the same Hamiltonian and angular momentum cutoff. The neural energies are systematically lower than the 2D DVR estimates, with discrepancies $\Delta_{\mathcal{E}}<1\%$ at most tested frequencies and a larger discrepancy $\Delta_{\mathcal{E}}\approx2.5\%$ at the lowest frequency $\nu=68.31\,\text{MHz}$. Since the accuracy of the trimer ${(N=3)}$ reference is less firmly established than that of the dimer MDVR benchmark, these discrepancies quantify agreement between the two solvers. The agreement demonstrates that our variational method can resolve coupled angular channels and correlated radial motion together in a field-linked three-body system.

%\begin{wrapfigure}{r}{0.5\textwidth}
% \vspace{-3em}
%    \centering
%    \includegraphics[width=0.5\textwidth]{figures/grid_comparison.pdf}
%\caption{
%\textbf{Effect of sampling strategy.} Training trajectories for ground and excited states using uniform, logarithmic, and WKB-based sampling. Uniform sampling leads to noisy optimization and slow convergence, while logarithmic and WKB-based sampling improve stability and convergence.}
%    \label{fig_grid}
    % \vspace{-1em}
%\end{wrapfigure}

%\subsection{Effect of sampling strategy}
%\label{sec_exp_sampling}

%We compare three sampling strategies: uniform, logarithmic, and the proposed WKB-based MCMC sampling. Uniform sampling leads to slow convergence and higher final energy due to poor coverage of relevant regions (Fig.~\ref{fig_grid}). Both logarithmic and WKB-based sampling improve optimization.
%While the logarithmic grid performs comparably in this setting, it relies on a fixed, problem-specific coordinate transformation. In contrast, the proposed approach adapts to the physical structure through the sampling distribution itself and does not require manual tuning. This distinction becomes critical in higher-dimensional settings, where suitable coordinate transformations are unavailable.

\section{Discussion}
\label{sec_conclusion}
% conclusion, limitations and outlook

In this work, we introduced a physics-informed neural framework for solving quantum few-body molecular problems with anisotropic long-range interactions. Rather than relying on handcrafted discretizations or problem-specific bases, the method combines a variational neural representation with symmetry-aware and asymptotic inductive biases derived directly from the Hamiltonian structure. Rotational symmetries are embedded analytically through spherical harmonic decomposition, while asymptotic behavior is incorporated through a tempered WKB-inspired envelope. Across interaction regimes, the approach accurately reproduces reference two-body MDVR solutions for both ground and excited states while maintaining stable optimization behavior. We further demonstrate the method on a field-linked three-body system,
simultaneously resolving coupled angular channels and correlated radial motion.  More broadly, these results suggest that embedding physical structure into neural wave function representations is a promising direction for scalable first-principles simulation of correlated molecular quantum systems and engineered quantum matter.

\hfill
\newpage

%While our primary benchmarks focus on two-body field-linked molecular systems to enable controlled evaluation against established baselines, we explicitly formulate the three-body extension and experimentally validate each of its core numerical components.

\subsection*{AI use statement}

% (This section is \textbf{required} and does not count toward the page limit.)

% In this work, we used generative AI tools for [tasks with required disclosure].
% We have not used generative AI tools for [other tasks with required disclosure],
% and [the rest of the required disclosure tasks] are not applicable to this work.
% Additionally, we used generative AI tools for [tasks with recommended
% disclosure]. We have reviewed all AI-assisted work. [Elaborate. For example, “we
% checked LLM-generated research ideas for potential plagiarism through a manual
% literature survey”, “LLM-generated code was verified and tested for correctness
% by 2 authors”, etc.]. We take responsibility for the final content of this work,
% including text, claims or artifacts produced with the aid of generative AI.

% In this work, we did not use generative AI tools for tasks with required disclosure.
We have not used generative AI tools for tasks with required disclosure.
% , and the rest of the required disclosure tasks are not applicable to this work.
Additionally, we used generative AI tools for editing text, unit testing and generating figures in code, and identifying the relevant literature. We have reviewed all AI-assisted work. We made sure the text is consistent with our research direction and accurate and the code runs correctly without any biases. We take responsibility for the final content of this work,
including text, claims or artifacts produced with the aid of generative AI.

% See the ICLR 2027 AI Policy for Authors for more details. This statement should
% not be more than 1 page.

% \subsection*{Ethics statement}

% (This section is \textbf{recommended} and does not count toward the page limit.)

% If authors feel that their paper submission raises questions regarding the Code
% of Ethics, they are encouraged to include a paragraph of Ethics Statement (at
% the end of the main text before references) to address potential concerns where
% appropriate. Topics include, but are not limited to, studies that involve human
% subjects, practices to data set releases, potentially harmful insights,
% methodologies and applications, potential conflicts of interest and sponsorship,
% discrimination/bias/fairness concerns, privacy and security issues, legal
% compliance, and research integrity issues (e.g., IRB, documentation, research
% ethics). This statement should not be more than 1 page.

\subsection*{Reproducibility statement}

To ensure full reproducibility of our results, all algorithmic details, hyperparameter configurations, and experimental setups are provided in the Appendix. Specifically, the complete optimization procedure and model workflow are summarized as pseudocode in App.~\ref{app_pseudo_code}, while training hyperparameters, network architectures, and technical details are documented in App.~\ref{app_technical_details}. Furthermore, to verify optimization stability against stochastic initialization, we report mean and standard deviation metrics across multiple random seeds in App.~\ref{app_exp_robustness}. Source code and data generation scripts will be made publicly available upon publication.
% (This section is \textbf{recommended} and does not count toward the page limit.)

% It is important that the work published in ICLR is reproducible. Authors are
% strongly encouraged to include a paragraph-long Reproducibility Statement at the
% end of the main text (before references) to discuss the efforts that have been
% made to ensure reproducibility. This paragraph should not itself describe
% details needed for reproducing the results, but rather reference the parts of
% the main paper, appendix, and supplemental materials that will help with
% reproducibility. For example, for novel models or algorithms, a link to an
% anonymous downloadable source code can be submitted as supplementary materials;
% for theoretical results, clear explanations of any assumptions and a complete
% proof of the claims can be included in the appendix; for any datasets used in
% the experiments, a complete description of the data processing steps can be
% provided in the supplementary materials. Each of the above are examples of
% things that can be referenced in the reproducibility statement.



\bibliography{ARXIV}
\bibliographystyle{iclr2027_conference}

\appendix
\newpage

\section*{\Large Appendix}

\section{Extended physical model}
\label{app_physical_model}

\subsection{Jacobi coordinates}
\label{app_Jacobi}

To simplify analytical expressions that describe few-body systems, it is useful to define a reference frame with origin at the system's \textit{center of mass (CM)}, whose coordinate and gradient are defined as
\begin{equation}
	\bm{R}_\text{CM}=\sum\limits_{j=1}^N \frac{m_j}{M}\bm{R}_j\quad\Rightarrow\quad\bm{\nabla}_{\bm{R}_\text{CM}}=\sum_{j=1}^N \bm{\nabla}_{\bm{R}_j}\,,
\end{equation}
where $M=\sum_{j=1}^N m_j$ is the total mass of the system, and $m_j$, $\bm{R}_j$ and $\bm{\nabla}_{\bm{R}_j}$ are respectively mass, coordinate and gradient of the $j$-th particle. The na\"ive definition of particle coordinates in the CM reference frame is
\begin{equation}
	\bm{\rho}_j = \bm{R}_j-\bm{R}_\text{CM}\,,
	\label{eq_CM_coordinates_naive}
\end{equation}
which provides a set of \textit{non-independent} vectors, since $\sum_j m_j\bm{\rho}_j=0$. The laboratory-frame $(\bm{\nabla}_{\bm{R}_j})$ and CM-frame~$(\bm{\nabla}_{\bm{\rho}_j})$ gradients are related via
\begin{equation}
	\bm{\nabla}_{\bm{R}_j}=\frac{m_j}{M}\bm{\nabla}_{\bm{R}_\text{CM}}+\bm{\nabla}_{\bm{\rho}_j}-\frac{m_j}{M}\left(\sum\nolimits_{k=1}^N \bm{\nabla}_{\bm{\rho}_k}\right)\,.
\end{equation}
The coordinate choice~\eqref{eq_CM_coordinates_naive} results in a convoluted expression for the kinetic energy~\eqref{eq_hamiltonian_few_body}
\begin{align}
	\hat{\mathcal{H}}_\text{kin} = \sum\limits_{j=1}^N\mathcal{E}^{\text{kin}}_{j}\bm{\nabla}^2_{\bm{R}_j} &=  \sum\limits_{j=1}^N\mathcal{E}^{\text{kin}}_{j}\left[\frac{m_j}{M}\bm{\nabla}_{\bm{R}_\text{CM}}+\bm{\nabla}_{\bm{\rho}_j}-\frac{m_j}{M}\bigg(\sum\nolimits_{k=1}^N \bm{\nabla}_{\bm{\rho}_k}\bigg)\right]^2\nonumber\\
	&=-\frac{\hbar^2}{2M}\bm{\nabla}^2_{\bm{R}_\text{CM}}+\sum\limits_{j=1}^N\mathcal{E}^{\text{kin}}_{j}\bm{\nabla}^2_{\bm{\rho}_j}+\frac{\hbar^2}{2M}\sum\limits_{j,k=1}^N \big(\bm{\nabla}_{\bm{\rho}_j}\cdot\bm{\nabla}_{\bm{\rho}_k}\big)\,,
\end{align}
where kinetic factors $\mathcal{E}^{\text{kin}}_{j}=-\hbar^2/(2m_j)$, and the mass polarization term, containing the scalar product, arises, since vectors~$\bm{\rho}_j$ are not independent.

The issue with the non-independence can be resolved by considering \textit{Jacobi coordinates}~$\bm{r}_j$~\citep{Suzuki_1998,Jacobi_Lectures}. These vectors are defined recursively for $j=1,...,N-1$, as
\begin{equation}
	\bm{r}_{j} = \frac{1}{M_j}\bigg(\sum\nolimits_{k=1}^j m_k\bm{R}_k\bigg) - \bm{R}_{j+1}\,,
    \label{eq_Jacobi_coordinates}
\end{equation}
where the Jacobi masses $M_j=\sum_{k=1}^j m_k$, and the last vector $\bm{r}_N\equiv\bm{R}_\text{CM}$. Intuitively, the Jacobi method recursively replaces a pair of particles with a virtual particle placed at their center of mass. The process is repeated, until one reaches the overall center of mass of an $N$-particle system.

The kinetic energy~\eqref{eq_hamiltonian_few_body} of the system can be expressed in Jacobi coordinates~\eqref{eq_Jacobi_coordinates} the following way:
\begin{align}
	\hat{\mathcal{H}}_\text{kin} 
	&=-\frac{\hbar^2}{2M}\bm{\nabla}^2_{\bm{R}_\text{CM}}-\sum\limits_{j=1}^{N-1}\frac{\hbar^2}{2\mu_j}\bm{\nabla}^2_{\bm{r}_j}\,,
	\label{eq_kin_en_CM}
\end{align}
where the reduced masses $\mu_j=m_{j+1}M_j/M_{j+1}$. Here, the first term is the kinetic energy of the center of mass, which is usually omitted, and the second term is the total kinetic energy of particles in the CM reference frame. Ultimately, the generic $N$-particle Hamiltonian can be approximated by
\begin{equation}
	\hat{\mathcal{H}}=-\sum\limits_{j=1}^{N-1}\frac{\hbar^2}{2\mu_j}\bm{\nabla}^2_{\bm{r}_j}+\sum\limits_{i<j}^{N} V(\tilde{\bm{R}}_{i}-\tilde{\bm{R}}_{j})\,,
\end{equation}
where $\tilde{\bm{R}}_{i,j}$ need to be treated as functions of Jacobi coordinates~$\bm{r}_{i}$.

Consequently, the wave function factorizes as
\begin{equation}
    \Psi(\bm{R}_1,...,\bm{R}_N) =
    \Psi_{\text{CM}}(\bm{R}_{\text{CM}})\,
    \Psi_{\text{internal}}(\bm{r}_1,...,\bm{r}_{N-1})\,.
\end{equation}
The center-of-mass factor describes free motion of the entire system and does
not affect the intrinsic bound-state structure. Discarding
$\bm{R}_{\text{CM}}$ is therefore equivalent to working in the intrinsic frame,
or quotienting out the global translation symmetry. In our PINN formulation,
the neural network represents only the internal wave function
$\Psi_{\text{internal}}(\bm{r}_1,...,\bm{r}_{N-1})$, reducing the input dimension from $3N$ to
$3(N-1)$ while enforcing translational invariance exactly.

\subsection{Microwave-shielded interaction potential}
\label{app_field_linked}

From the physics perspective, we are interested in ultracold diatomic molecules, hereafter referred to as monomers, that have a large permanent electric dipole moment $d$. The dipole moment is measured in the body-fixed frame, but it averages to zero in the laboratory frame. To get a net dipole moment in the laboratory frame, one can couple the transition (with frequency $\omega_{0}$) from the ground to the first excited rotational state of the monomer, to a microwave field (with frequency  $\omega$). The coupling strength is characterized by the Rabi frequency $\Omega=2\pi\nu$, defined as the product of the dipole moment~$d$ and the amplitude of the microwave electric field $F_\text{mw}$. For convenience, we also define here the rescaled detuning $\delta=(\omega_{0}-\omega)/\Omega$ by the Rabi frequency. Not only does the microwave-field-to-molecule coupling induce a net dipole moment $d_{\rm eff}=d/\sqrt{12 (1+\delta^2)}$ in the dressed monomer state, but also—more importantly—a repulsive core in the dipole–dipole interaction potential between two monomers. This, so-called, microwave shielding \citep{Karman2018} is crucial for maintaining the stability of ultracold dipolar molecules, which would otherwise undergo significant losses when they encounter each other at short range. The adiabatic shielding potential between two dressed monomers can be given in a simple closed form~\citep{Deng_2023}
\begin{equation}
    V(\bm{r})=C_6 [1-\cos^4{\theta}]\,r^{-6} + C_3[3\cos^2 \theta-1]\,r^{-3}\,,
    \label{eq_V_eff}
\end{equation}
where $C_3=d_{\rm eff}^2/(4 \pi \epsilon_0)=d^2/[48 \pi \epsilon_0 (1+\delta^2)]$, $C_6=d^4/[128 \pi^2 \epsilon_0^2 \hbar\Omega (1+\delta^2)^{3/2}]$ and $\bm{r}=\{r,\theta,\phi\}$ is the relative interparticle  in spherical coordinates. Note that the potential can also depend on $\phi$, when one turns on the ellipticity $\xi \ne 0$ of the field~\citep{Deng_2023}. The sign of $\delta$ plays no role in Eq.~\eqref{eq_V_eff}, however, in practice, we do need a minus sign (blue detuning) for it. In our study, we can simply set the microwave field to be on resonance: $\delta=0$.

For numerical calculations, it is instrumental to use the \textit{atomic unit (a.u.) system}, in which $\hbar=1$ and $4 \pi \epsilon_0=1$. Using these units, the interaction potential reads
\begin{equation}
	V^{\text{a.u.}}_{\text{eff}}(\bm{r})=C_6^{\rm a.u.} [1-\cos^4\theta]\,r^{-6}+C_3^{\rm a.u.}[3\cos^2\theta-1]\,r^{-3},
    \label{eq_interaction_potential}
\end{equation}
where $C_6^{\rm a.u.}= (d^{\rm a.u.})^4/(8 \Omega^{a.u.})$ and $C_3^{\rm a.u.}= (d^{\rm a.u.})^2/12$. Note, that we already set $\delta=0$. \mbox{The actual physical parameters} for $^{23}$Na$^{40}$K in SI units: 
\begin{equation*}
    m=62.95376748\,\text{Da}\,,
    \qquad
    d=2.72\,\text{D}\,,
    \qquad
    \Omega/(2\pi)=\nu\in[16,50]\,\text{MHz}\,.
\end{equation*}
Conversion from SI units to a.u.: 
\begin{equation*}
    [1\,\text{Da}]_{\text{a.u.}}=1822.888486\,[m_e]_{\text{a.u.}}\,,
    \qquad
    [1\,\text{D}]_{\text{a.u.}}=0.3934303\,[ea_0]_{\text{a.u.}}\,,
\end{equation*}
\begin{equation*}
    [1\,\text{MHz}]_{\text{a.u.}}=1.5198298460570\times 10^{-10}\,[E/h]_{\text{a.u.}}\,.
\end{equation*}

\subsection{Two-body system}
\label{app_two_body}

For a system with $N=2$ molecules, there is a single Jacobi coordinate (see App.~\ref{app_Jacobi}). Namely, the relative interparticle position: $\bm{r}=\bm{R}_1-\bm{R}_2$. In this setting, the Hamiltonian~\eqref{eq_hamiltonian_few_body} reads
\begin{equation}
	\hat{\mathcal{H}}=-\frac{\hbar^2}{2\mu}\bm{\nabla}^2_{\bm{r}}+V(\bm{r})\,,
	\label{eq_2B_hamiltonian}
\end{equation}
where $\mu=m_1 m_2/(m_1+m_2)=m/2$ is the reduced mass.

\subsubsection{Cylindrical coordinates}
\label{app_two_body_cylindrical}

Since the interaction potential~\eqref{eq_interaction_potential} does not depend on the spherical angle $\phi$, the straightforward way to solve the Hamiltonian~\eqref{eq_2B_hamiltonian} is by using reduced cylindrical coordinates $\bm{r}=\{r_\perp,z\}$ with the angle~$\phi$ omitted. In particular, one considers the following wave function
\begin{equation}
    \Psi(\bm{r})=\psi_{\text{2D}}(r_{\perp},z)/\sqrt{r_{\perp}}\,,
    \label{eq_app_cylindrical_wavefunction}
\end{equation}
for which the kinetic energy contribution to the Hamiltonian reads
\begin{equation}
    \hat{\mathcal{H}}_{\text{kin}}=-\frac{\hbar^2}{2\mu}\bm{\nabla}^2_{\bm{r}}=-\frac{\hbar^2}{2\mu}\left(\bm{\nabla}^2_{\bm{r_\perp}}+\bm{\nabla}^2_{\bm{z}}+\frac{1}{4r_{\perp}^2}\right)\,,
\end{equation}
and the potential can be obtained by substituting into Eq.~\eqref{eq_interaction_potential}
\begin{equation}
    r=\sqrt{r_{\perp}^2+z^2}\quad\text{and}\quad \cos\theta=z/r\,.
\end{equation}
The design of the optimization procedure for wave function~\eqref{eq_app_cylindrical_wavefunction} is thoroughly explained in App.~\ref{app_2d_explanations}.

\subsubsection{Spherical decomposition}
\label{app_spherical_decomposition}

To simplify the learning problem, one can analytically take care of the angular behavior of the wave function. In particular, wave function $\Psi(\bm{r})$ and the interaction potential $V(\bm{r})$ can be expanded over the complete basis of spherical harmonics:
\begin{gather}
    \Psi(\bm{r})=\sum\limits_{l,m}\psi_{l,m}(r)\,Y_{l,m}(\theta,\phi)/r
    \quad\text{and}\quad
	V(\bm{r})=\sum\limits_{l,m}V_{l,m}(r)\,Y_{l,m}(\theta,\phi)\,,
    \label{eq_2B_sph_exp_V}
\end{gather}
where $\bm{r}=\{r,\theta,\phi\}$ are spherical coordinates and $Y_{l,m}(\theta,\phi)$, where  $l=0,1,2,...$ and $m\in\{-l,l\}$, are spherical harmonics with the Condon-Shortley phase~\citep{Varshalovich_Book}. The spherical components $V_{l,m}(r)$ of the interaction potential~\eqref{eq_potential} can be explicitly calculated
\begin{equation}
    V_{0,0}(r) = \frac{8\sqrt{\pi}C_6}{5}\frac{1}{r^6}
    \,\,\text{,}\,\,
    V_{2,0}(r) = 4\sqrt{\frac{\pi}{5}}\bigg(C_3\frac{1}{r^3}-\frac{2C_6}{7}\frac{1}{r^6}\bigg)
    \,\,\text{,}\,\,
    V_{4,0}(r) = -\frac{16\sqrt{\pi}C_6}{105}\frac{1}{r^6}\,.
\end{equation}

Expansions~\eqref{eq_2B_sph_exp_V} allow us to remove angles $\{\theta,\phi\}$ from the numerical optimization problem. After applying these expansions, the energy functional~\eqref{eq_energy_functional} can be written down as
\begin{equation}
    \mathcal{E}_{\bm{w}}[\hat{\mathcal{H}}]\sim
    \int\Psi_{\bm{w}}^{\ast}(\bm{r})\,\hat{\mathcal{H}}\,\Psi_{\bm{w}}(\bm{r})\mathrm{d}\bm{r}
    =\int\sum\limits_{l,m;l',m'}\psi^{\ast}_{\bm{w},l',m'}(r)\hat{\mathcal{H}}_{l,m;l',m'}(r)\psi_{\bm{w},l,m}(r)\,\mathrm{d}r\,,
    \label{eq_2B_sph_exp_energy_functional}
\end{equation}
where $\bm{w}$ are neural network parameters, and the cross-channel Hamiltonian
\begin{equation}
	\hat{\mathcal{H}}_{l,m;l',m'}(r)\equiv\hat{\mathcal{H}}^\text{kin}_{l,m;l',m'}(r)+\mathcal{H}^\text{pot}_{l,m;l',m'}(r)\,.
\end{equation}
The cross-channel potential Hamiltonian reads
\begin{equation}
	\mathcal{H}^\text{pot}_{l,m;l',m'}(r)\equiv \sum\limits_{L,M}V_{L,M}(r)G^{L,M}_{l,m;l',m'}\,,
	\label{eq_2B_potential}
\end{equation}
where the Gaunt coefficients are defined as integrals
\begin{equation}
    G^{L,M}_{l,m;l',m'} = \int Y^{\ast}_{l',m'}(\Omega) Y_{L,M}(\Omega) Y_{l,m}(\Omega)\,\mathrm{d}\Omega
    =\sqrt{\frac{(2l+1)(2L+1)}{4\pi(2l'+1)}}C^{l',m'}_{l,m;L,M}C^{l',0}_{l,0;L,0}\,,
    \label{eq_Gaunt}
\end{equation}
where $C^{l',m'}_{l,m;L,M}$ are the Clebsch--Gordan coefficients that encapsulate angular-momentum algebra. The cross-channel kinetic Hamiltonian is defined as
\begin{equation}
	\hat{\mathcal{H}}^\text{kin}_{l,m;l',m'}(r)=\hat{\mathcal{H}}^{\text{kin},\text{rad}}_{l,m;l',m'}+\mathcal{H}^{\text{kin},\text{cent}}_{l,m;l',m'}(r)\,,
	\label{eq_2B_kinetic}
\end{equation}
with the diagonal radial derivative and centrifugal terms defined as
\begin{equation}
	\hat{\mathcal{H}}^{\text{kin},\text{rad}}_{l,m;l',m'}=-\delta_{l,l'}\delta_{m,m'}\frac{\hbar^2}{2\mu}\frac{\partial^2}{\partial r^2}
    \quad\text{and}\quad
    \mathcal{H}^{\text{kin},\text{cent}}_{l,m;l',m'}(r)= \delta_{l,l'}\delta_{m,m'}\frac{\hbar^2}{2\mu}\frac{l(l+1)}{r^2}\,,
\end{equation}
where $\delta_{l,l'}$ is the Kronecker delta. One has to be careful, when implementing the operator $\hat{\mathcal{H}}^{\text{kin},\text{rad}}_{l,m;l',m'}$ within a neural network. We integrate the radial kinetic-energy expectation by parts, with vanishing boundary terms, to obtain a positive gradient-squared form suitable for numerical optimization.

\subsection{Three-body system}
\label{app_three_body}

Here, we derive the radial Hamiltonian matrix elements used in the three-body calculations of Sec.~\ref{sec_three_body}.

For a system of $N=3$ particles of the same mass $m$ at positions $\{\bm{R}_1, \bm{R}_2, \bm{R}_3\}$, the Jacobi coordinates~(see App.~\ref{app_Jacobi}) are defined as 
\begin{equation}
    \bm{r}_{1}=\bm{R}_1-\bm{R}_2\,,
    \qquad
    \bm{r}_{2}=(\bm{R}_1+\bm{R}_2)/2-\bm{R}_3\,,
    \qquad
    \bm{r}_3=\bm{R}_{\text{CM}}=(\bm{R}_1+\bm{R}_2+\bm{R}_3)/3\,.
\end{equation}
The pair-separation vectors can be expressed as
\begin{equation}
    \bm{R}_1-\bm{R}_2=\bm{r}_{1}\,,
    \qquad
    \bm{R}_2-\bm{R}_3=\bm{r}_2-\bm{r}_{1}/2\,,
    \qquad
    \bm{R}_3-\bm{R}_1=-\bm{r}_2-\bm{r}_{1}/2\,. 
\end{equation}
In the absence of an external potential, the free center-of-mass motion that depends on $\bm{R}_{\text{CM}}$ can be trivially described by a plane wave, leaving only the relative degrees of freedom in the Hamiltonian:
\begin{equation}
    \hat{\mathcal{H}}_{\text{3-body}}
=
-\frac{\hbar^2}{m}\nabla_{\bm{r}_1}^{2}
-\frac{3\hbar^2}{4m}\nabla_{\bm {r}_2}^{2}
+V(\bm{r}_1)
+V\!\left(\bm{r}_2-\bm{r}_1/2\right)
+V\!\left(-\bm{r}_2-\bm{r}_1/2\right).
\end{equation}
We implement the spherical harmonic expansion~(see App.~\ref{app_spherical_decomposition}) of the wave function $\Psi_{\text{3-body}}(\bm{r}_1,\bm{r}_2)$ on both $\bm{r}_1$ and $\bm{r}_2$:
\begin{equation}
    \Psi_{\text{3-body}}(\bm{r}_1,\bm{r}_2)
    =
    \frac{1}{r_1r_2}
    \sum_{l,m_l;L,M_L}
    \psi^{\text{2D}}_{l,m_l;L,M_L}(r_1,r_2)\,
    \big[Y_{l,m_l}(\theta_{1},\phi_{1})\otimes
    Y_{L,M_L}(\theta_{2},\phi_{2})\big]\,,
    \label{eq_3B_expansion}
\end{equation}
where $\bm{r}_{1}=\{r_{1},\theta_{1},\phi_{1}\}$ and $\bm{r}_{2}=\{r_{2},\theta_{2},\phi_{2}\}$ are the spherical coordinates, $\psi^{\text{2D}}_{l,m_l;L,M_L}(r_1,r_2)$ is the 2D radial
wave function discussed in App.~\ref{app_2d_explanations}, and
the term in square brackets is the tensor product of spherical
harmonics (see~\citet{Varshalovich_Book}).

As in the two-body case, the purpose of
expansion~\eqref{eq_3B_expansion} is to eliminate the angular
coordinates, leaving only the two radial coordinates $\{r_1,r_2\}$
for the numerical optimization. The single-coordinate interaction
$V(\bm r_1)$ is treated as in App.~\ref{app_spherical_decomposition}.
The remaining pair interactions depend on linear combinations of
the two Jacobi coordinates and therefore couple their angular
momenta. The resulting Hamiltonian matrix elements read
\begin{align}
    &\mathcal H_{l,m_l;L,M_L\to l',m'_l;L',M'_L}(r_1,r_2)
    =\nonumber\\
    &\hspace{4em}\mathcal H^{\text{kin}}_{l,L}(r_1,r_2)
    \delta_{l,l'}\delta_{m_l,m'_l}
    \delta_{L,L'}\delta_{M_L,M'_L}
    +
    V_{l,m_l;l',m'_l}(r_1)\delta_{L,L'}\delta_{M_L,M'_L}
    \nonumber\\
    &\hspace{8em}+
    V^{(23)}_{l,m_l;L,M_L\to l',m'_l;L',M'_L}(r_1,r_2)
    +
    V^{(31)}_{l,m_l;L,M_L\to l',m'_l;L',M'_L}(r_1,r_2)\,,
    \label{eq_3B_hamiltonian}
\end{align}
where $V_{l,m_l;l',m'_l}(r_1)\equiv
\mathcal H^{\text{pot}}_{l,m_l;l',m'_l}(r_1)$ from
Eq.~\eqref{eq_2B_potential}, and the kinetic Hamiltonian is
\begin{equation}
    \mathcal H^{\text{kin}}_{l,L}(r_1,r_2)
    =
    -\frac{\hbar^2}{m}
    \left(
        \frac{\partial^2}{\partial r_1^2}
        -\frac{l(l+1)}{r_1^2}
    \right)
    -\frac{3\hbar^2}{4m}
    \left(
        \frac{\partial^2}{\partial r_2^2}
        -\frac{L(L+1)}{r_2^2}
    \right)\,.
\end{equation}

To evaluate the remaining pair interactions, we first consider the
pair $\{\bm{R}_{2},\bm{R}_{3}\}$, with separation vector
$\bm\rho_{23}=\bm r_2-\bm r_1/2$. Using the spherical components
$V_{l,m}(r)$ defined in App.~\ref{app_spherical_decomposition},
we write
\begin{equation}
    V(\bm\rho_{23})
    =
    \sum_{\lambda\in\{0,2,4\}}
    V_{\lambda,0}(\rho_{23})
    Y_{\lambda,0}(\theta_{23},\phi_{23})\,,
    \label{eq_3B_potential_separation}
\end{equation}
where $\bm\rho_{23}=\{\rho_{23},\theta_{23},\phi_{23}\}$ are the spherical coordinates.
We expand each spherical harmonic in Eq.~\eqref{eq_3B_potential_separation} into a product of spherical harmonics of Jacobi coordinates $\{\bm{r}_{1},\bm{r}_2\}$, using the solid harmonic expansion. Denoting the angular ranks
on $\bm r_2$ and $\bm r_1$ by $a$ and $b$, respectively, we obtain
\begin{align}
    Y_{\lambda,0}(\theta_{23},\phi_{23})
    =\rho_{23}^{-\lambda}
    \sum_{\substack{a,b\geq 0\\a+b=\lambda}}
    (-1)^b r_2^a & \left(\frac{r_1}{2}\right)^b
    \sqrt{\frac{4\pi(2\lambda+1)!}{(2a+1)!(2b+1)!}}
    \nonumber\\
    &\quad\times
    \sum_{m_a+m_b=0}
    C^{\lambda,0}_{a,m_a;b,m_b}
    Y_{a,m_a}(\theta_{2},\phi_{2})
    Y_{b,m_b}(\theta_{1},\phi_{1})\,.
    \label{eq_3B_potential_expansion}
\end{align}
For the pair $\{\bm{R}_{3},\bm{R}_{1}\}$, the separation vector is
$\bm\rho_{31}=-\bm r_2-\bm r_1/2$. Both Jacobi vectors enter with
a minus sign, so the phase in Eq.~\eqref{eq_3B_potential_expansion}
becomes $(-1)^{a+b}=(-1)^\lambda$.

After expansion~\eqref{eq_3B_potential_expansion}, the remaining
scalar factor still depends on the relative angle between $\bm{r}_{1}$ and $\bm{r}_{2}$ through
$\xi=(\bm r_1\cdot\bm r_2)/(r_{1}r_{2})$. In particular, the two pair distances can be expanded as
\begin{equation}
    \rho_{23}(\xi)=\sqrt{r_2^2+r_1^2/4-r_1r_2\xi}\,,
    \qquad
    \rho_{31}(\xi)=\sqrt{r_2^2+r_1^2/4+r_1r_2\xi}\,.
\end{equation}
For either pair $p\in\{23,31\}$, we expand this factor as
\begin{equation}
    V_{\lambda,0}\big(\rho_p(\xi)\big) \cdot \rho_{p}(\xi)^{-\lambda}
    =
    \sum_{\kappa=0}^{\infty}\sum_{\nu=-\kappa}^{\kappa}
    \mathcal F^{\kappa}_{\lambda,p}(r_1,r_2)
    Y_{\kappa,\nu}(\theta_{2},\phi_{2})
    Y_{\kappa,\nu}^{*}(\theta_{1},\phi_{1})\,,
\end{equation}
where the radial coefficients are given by
\begin{equation}
    \mathcal F^{\kappa}_{\lambda,p}(r_1,r_2)
    =
    2\pi\int_{-1}^{1}
    V_{\lambda,0}\big(\rho_p(\xi)\big)\rho_p(\xi)^{-\lambda}
    P_{\kappa}(\xi)\,\mathrm{d}\xi\,,
\end{equation}
and $P_\kappa(\xi)$ is the Legendre polynomial of degree $\kappa$.

The angular integrals over $\{\theta_{1},\phi_{1}\}$ and $\{\theta_{2},\phi_{2}\}$
can now be evaluated separately. Using the Gaunt coefficients
defined in Eq.~\eqref{eq_Gaunt}, we write each integral as
\begin{align}
    \mathcal I^{J,M;J',M'}_{a,m_a;\kappa,q}
    &\equiv
    \int
    Y_{J,M}^{*}(\Omega)
    Y_{a,m_a}(\Omega)
    Y_{\kappa,q}(\Omega)
    Y_{J',M'}(\Omega)\,\mathrm{d}\Omega
=
    \sum_j
    G^{a,m_a}_{\kappa,q;j,m_a+q}
    G^{j,m_a+q}_{J',M';J,M}\,,
\end{align}
where $\Omega\equiv\{\theta,\phi\}$ and $\mathrm{d}\Omega\equiv\sin\theta\,\mathrm{d}\theta\,\mathrm{d}\phi$. The matrix elements of the two pair interactions then read
\begin{align}
    &V^{(p)}_{l,m_l;L,M_L\to l',m'_l;L',M'_L}(r_1,r_2)
    \nonumber\\
    &\hspace{2em}=
    \sum_{\lambda\in\{0,2,4\}}
    \sum_{\substack{a,b\geq 0\\a+b=\lambda}}
    s_p(a,b)\,r_2^a\left(\frac{r_1}{2}\right)^b
    \sqrt{\frac{4\pi(2\lambda+1)!}{(2a+1)!(2b+1)!}}
    \nonumber\\
    &\hspace{4em}\times
    \sum_{m_a+m_b=0}
    C^{\lambda,0}_{a,m_a;b,m_b}
    \sum_{\kappa=0}^{\infty}\sum_{\nu=-\kappa}^{\kappa}
    \mathcal F^{\kappa}_{\lambda,p}(r_1,r_2)(-1)^\nu
    \mathcal I^{L,M_L;L',M'_L}_{a,m_a;\kappa,\nu}
    \mathcal I^{l,m_l;l',m'_l}_{b,m_b;\kappa,-\nu}\,,
\end{align}
where $s_{23}(a,b)=(-1)^b$ and $s_{31}(a,b)=(-1)^{a+b}$.

In summary, expansion~\eqref{eq_3B_expansion} eliminates angular
coordinates from the three-body optimization problem. The
Clebsch--Gordan and Gaunt coefficients encode the angular-momentum
coupling between the two Jacobi coordinates, leaving a matrix of
radial operators in Eq.~\eqref{eq_3B_hamiltonian}. The neural
representation therefore only needs to learn the corresponding
two-dimensional radial channel functions
$\psi^{\text{2D}}_{l,m_l;L,M_L}(r_1,r_2)$. The same construction
generalizes to additional Jacobi coordinates through higher tensor
products of spherical harmonics and standard angular-momentum
coupling algebra.

%\textcolor{red}{FORMATTING ABOVE!}

\section{Extended method}
\label{app_extended_method}

\subsection{Properties of the Spherical Harmonic Representation}
\label{app_spherical_motivation}
This section summarizes several mathematical and physical properties of spherical harmonic representations that motivate their use in the proposed neural wave function parametrization. While not required for the formulation itself, these properties explain the numerical stability, scalability, and physical interpretability of the decomposition.

\textbf{I. The natural basis.} From the mathematical perspective, spherical harmonics $Y_{l,m}(\theta,\phi)$ form a complete orthonormal basis for $L^2$-integrable functions of angles $\{\theta,\phi\}$. Unlike other polynomial sets, like Hermite or Chebyshev polynomials, commonly used for numerical methods, the use of spherical harmonics is not motivated by the boundary conditions of the problem. Problem-specific functional bases are a common cause of the fragility of numerical methods: changing the interaction or boundary conditions would require redesigning the representation. In contrast, for every value of $l$, spherical harmonics form the standard \textit{irreducible} representation of the SO$(3)$ group and provide the natural, symmetry-inspired basis for functions of angles $\{\theta,\phi\}$.

\textbf{II. Dealing with singularities.} Spherical coordinates $\bm{r}=\{r,\theta,\phi\}$ are curvilinear. Because of that, they are singular at the origin $(r=0)$ and along the polar axes $(\theta\in\{0,\pi\})$. Differential operators, such as the Laplacian $\bm{\nabla}^2_{\bm{r}}$ that constitutes kinetic energy~\eqref{eq_hamiltonian_few_body}, include terms, like $1/\sin{\theta}$, that behave numerically unstable around singular points. This highly affects the performance of numerical methods that employ spherical coordinates. In contrast, spherical harmonics $Y_{l,m}(\theta,\phi)$ embed singularities of the coordinate system into their structure. Thus, they provide numerically stable expressions for differential operators, like
\begin{equation}
    \bm{\nabla}^2_{\bm{r}} Y_{l,m}(\theta,\phi)\sim -l(l+1)\,Y_{l,m}(\theta,\phi)/r^2\,.
    \label{eq_centrifugal_term}
\end{equation}
The expectation value of the kinetic energy~\eqref{eq_hamiltonian_few_body} in terms of angular momentum pairs~$\{l,m\}$ is diagonal: $\langle Y_{l',m'}|\hat{\mathcal{H}}_{\text{kin}}|Y_{l,m}\rangle\sim\delta_{l,l'}\delta_{m,m'}$, which substantially reduces the computational cost.

\textbf{III. Scalability.} For an $N$-body system, the angular dependence of $N-1$ relative coordinates can be represented by tensor products of spherical harmonics. Under rotations, these products transform according to tensor-product representations of SO$(3)$, which can be decomposed into total-angular-momentum sectors using standard Clebsch--Gordan coupling. Consequently, angular-momentum coupling and transfer between different relative coordinates are handled analytically through Clebsch--Gordan and Wigner coefficients, rather than evaluated numerically. This construction extends recursively to an arbitrary number of coordinates, allowing the rotational structure of the full $N$-body wave function to be tracked algebraically. For example, a three-body wave function $\Psi_{\text{3-body}}(\bm{r_1},\bm{r_2})$ can be expanded in coupled products of spherical harmonics~(see Eq.~\eqref{eq_3B_wavefunction} and App.~\ref{app_three_body}).

\textbf{IV. Truncation.} Truncating the expansion $ {\Psi}(\bm{r}) = \sum_{l,m} \psi_{l,m}(r)\,Y_{l,m}(\theta,\phi)/r,$ at a finite $l_{\text{max}}$ is physically justified. Higher-$l$ harmonics $Y_{l,m}(\theta,\phi)$ are associated with larger centrifugal terms~\eqref{eq_centrifugal_term} in the expression for the kinetic energy~\eqref{eq_hamiltonian_few_body}. Similar to the interaction potential~\eqref{eq_potential}, these terms represent barriers to short-range molecular collisions. A higher barrier decreases the likelihood of molecule scattering into the higher-$l$ state, which renders these modes insignificant.

\subsection{Training pseudo-code}
\label{app_pseudo_code}

Algorithm~\ref{alg_training} summarizes the optimization procedure used throughout this work.
\begin{algorithm}[t]
\caption{Variational training with residual learning}
\label{alg_training}
\begin{algorithmic}[1]

\Require Hamiltonian $\hat{\mathcal{H}}$, WKB envelope $\psi_{\mathrm{WKB}}(r)$, angular cutoff $l_{\max}$, optimizer $\mathcal{O}$, reference energy $\mathcal{E}_{\mathrm{ref}}$
\State Initialize neural network parameters $\bm{\theta}$
\State Construct fixed MCMC sampling grid $\mathcal{G}=\{\bm{r}^{(i)}\}_{i=1}^{N}$ from the WKB-inspired density

\While{not converged}

    \State Sample mini-batch $\mathcal{B}\subset\mathcal{G}$

    \State Encode radial coordinates
    \[
    z^{(i)} = \mathrm{PE}(r^{(i)})
    \]

    \State Evaluate neural residual channels
    \[
    \psi^{\mathrm{NN}}_{l,m}(r^{(i)};\bm{\theta})
    =
    f_{\bm{\theta}}(z^{(i)})_{l,m}
    \]

    \State Construct radial functions
    \[
    \psi_{l,m}(r^{(i)})
    =
    \psi^{\mathrm{NN}}_{l,m}(r^{(i)};\bm{\theta})
    \psi_{\mathrm{WKB}}(r^{(i)})
    \]

    \State Assemble wave function
    \[
    \Psi(\bm{r}^{(i)})
    =
    \sum_{l,m}
    \psi_{l,m}(r^{(i)})
    Y_{l,m}(\theta^{(i)},\phi^{(i)})/r^{(i)}
    \]

    \State Evaluate normalized variational energy
    \[
    \mathcal{L}_{\bm{\theta}}
    =
    \frac{1}{|\mathcal{E}_{\mathrm{ref}}|}
    \frac{
    \langle \Psi_{\bm{\theta}}
    |
    \hat{\mathcal{H}}
    |
    \Psi_{\bm{\theta}}
    \rangle
    }{
    \langle
    \Psi_{\bm{\theta}}
    |
    \Psi_{\bm{\theta}}
    \rangle
    }
    \]

    \State Update parameters
    \[
    \bm{\theta}
    \leftarrow
    \mathcal{O}
    \left(
    \bm{\theta},
     \nabla_{\bm{\theta}}\mathcal{L}_{\bm{\theta}}
    \right)
    \]

\EndWhile

\end{algorithmic}
\end{algorithm}

\section{Validation baselines}
\label{app_dvr}

The discrete variable representation (DVR) is a standard numerical approach for solving the Schr\"odinger equation, in which the wave function is represented on a discrete grid and local potentials are diagonal in the grid indices. To benchmark our physics-informed neural network (PINN) for two-molecule systems, we extend the mapped discrete variable representation (MDVR)~\citep{Willner2004,Secker2021} to include multiple partial-wave couplings.

The MDVR approach defines a mapping function:
$x(r)=\frac{\sqrt{m}}{p_{\rm max}}\int_{r_{\rm in}}^{r}dr'\sqrt{E_{\rm max}-V_{\rm env}(r')}$,
such that a uniformly spaced grid in the $x$-representation corresponds to a denser $r$-grid where the local de Broglie wavelength, proportional to $1/\sqrt{E-V(r)}$, is small, and a coarser $r$-grid where it is large. Here $E_{\rm max}$ denotes the maximum energy considered, and $p_{\rm max}$ is the corresponding maximum local momentum. The envelope potential $V_{\rm env}(r)$ is set to the lowest adiabatic potential obtained by diagonalizing
$V^{\rm eff}_{ll'}(r)+\hbar^2 l(l+1)\delta_{ll'}/(2\mu r^2)$.
After the mapping, the remaining procedure in the $x$-representation is standard:
\begin{enumerate}[leftmargin=2.em]
    \setlength\itemsep{0em}
    \item Construct the Fourier DVR basis set $\{\tilde{s}_k(x)\}_{k=1}^{N-1}$ from uniformly spaced $x$-grid points $\{x_k\}_{k=0}^{N}$. The two endpoints are removed to impose hard-wall boundary conditions.
    \item Build the Hamiltonian matrix in this basis set:
    \[
    H_{il,jl'}=T_{ij}\delta_{ll'}
    +\left[
    V^{\rm eff}_{ll'}[r(x_i)]
    +\frac{\hbar^2 l(l+1)}{2\mu r(x_i)^2}\delta_{ll'}
    \right]\delta_{ij}.
    \]
    \item Diagonalize the Hamiltonian matrix to obtain the wave function in the $x$-representation:
    $\varphi_{n,l}(x)=\sum_i c^{i,l}_{n}\tilde{s}_i(x)/\sqrt{\Delta x}$,
    where $c^{i,l}_{n}$ is taken from the $n$-th eigenvector of the Hamiltonian matrix and $\Delta x=x_i-x_{i-1}$.
    \item Map the $x$-representation wave function back to the $r$-representation via
    $\psi_{n,l}(r)=\varphi_{n,l}[x(r)]/\sqrt{J[x(r)]}$,
    where $J(x)=dr(x)/dx$ denotes the Jacobian.
\end{enumerate}

We refer the reader to \citet{Willner2004} for the explicit expressions of the Fourier DVR basis $\{\tilde{s}_k(x)\}$ and the kinetic matrix elements $T_{ij}$. The MDVR can significantly reduce the numerical cost compared to DVR using uniformly spaced grids in the $r$-representation~\citep{Willner2004}. We use $N=512$ and $l\leq 8$ for two-molecule calculations. For comparison, a uniform-grid DVR with an illustrative spacing of $10\,a_0$ would require approximately $10^4$ radial points over the same domain. With five even partial waves, this gives a basis size of approximately $50000$, requiring 20~GB to store a fully dense, real double-precision Hamiltonian, excluding diagonalization workspace.

For three bodies, however, extending the radial mapping is more difficult because the other pair separations, $|\bm{r}_{2}\pm\bm{r}_{1}/2|$, can remain small even when both Jacobi radii are large. Coarsening the radial grids at large distances can therefore miss short-range pair correlations. A uniform-grid DVR with the same radial domain and spacing as in the two-body estimate, applied to both Jacobi radii, would require approximately $10^8$ radial grid pairs. With the 190-channel angular basis used in our refined three-body DVR calculations, the total basis size would reach approximately $1.9\cdot10^{10}$. A single real double-precision wave function vector would require approximately 152~GB, while storing the fully dense Hamiltonian would require approximately $2.89\cdot10^{21}$ bytes, making dense storage impractical. These figures illustrate the storage requirements of this uniform-grid DVR setup. Our current three-body DVR reference calculations focus on tightly bound trimers, using box boundaries of order $10^3\,a_0$ in each Jacobi radius and 50 grid points per coordinate.

These illustrative storage estimates motivate exploring neural radial-channel representations that avoid explicit storage of the full radial tensor-product grid. They are not convergence benchmarks: specialized classical methods can reduce this burden, including through structured or matrix-free Hamiltonian representations, and a computational advantage for the three-body neural approach remains to be demonstrated.

\section{Technical Details of the 2D Wave Function Representation}
\label{app_2d_explanations}

This Appendix details the two-dimensional wave function formulation used for the two-body problem in cylindrical coordinates and the three-body problem, including its numerical integration, differentiation, asymptotic envelope, and optimization.

\subsection{Integration and spatial derivatives}
\label{app_2d_integration}

We consider a two-dimensional wave function representation: either a two-body wave function~\eqref{eq_app_cylindrical_wavefunction} in reduced cylindrical coordinates~$ \{r_\perp,z\}$, or a three-body radial wave function~\eqref{eq_3B_expansion} in Jacobi coordinates~$\{r_1,r_{2}\}$. For generality, here we denote two coordinates as a single relative position vector $\bm{r}=\{r_1,r_{2}\}$, with $r=\sqrt{r_{1}^2+r_{2}^2}$. Both wave functions~\eqref{eq_app_cylindrical_wavefunction} and~\eqref{eq_3B_expansion} absorb the Jacobian into a radial pre-factor, giving
\begin{equation}
    \langle\Psi|\Psi\rangle
    =\gamma_{\text{sys}}\int_0^\infty\int_0^\infty
    |\psi_{\mathrm{2D}}(r_{1},r_{2})|^2\,\mathrm{d}r_{1}\,\mathrm{d}r_{2}\,,
    \label{eq_2d_norm}
\end{equation}
where $\gamma=1$ for the three-body system, and $\gamma=4\pi$ for the cylindrical two-body problem.

\paragraph{Sampling.}
We construct the 2D sampling grid from the one-dimensional coordinate
pool used in the radial formulation (Sec.~\ref{sec_sampling}).
At each optimization step, coordinates are sampled independently from
this pool and randomly paired to form points $\{r_{1,i},r_{2,i}\}$, yielding an
unstructured 2D point cloud. Points outside the prescribed radial
domain $r_{\min}\leq r\leq r_{\max}$ are discarded.

\paragraph{Triangulation and integration.}
We construct a Delaunay triangulation $\mathcal{T}$~\citep{Delaunay1934} of each retained
point cloud. For a triangle $T\in \mathcal{T}$ with vertices
$\{\bm{q}^{T}_{1},\bm{q}^{T}_{2},\bm{q}^{T}_{3}\}$, its area is
\begin{equation}
    A_T=\frac{1}{2}
    \left|\det\left(
    \bm{q}^{T}_{2}-\bm{q}^{T}_{1},
    \bm{q}^{T}_{3}-\bm{q}^{T}_{1}\right)\right|\,.
\end{equation}
We approximate integrals using the vertex quadrature
\begin{equation}
    \mathcal{I}_{\mathcal{T}}[f]
    =\gamma_{\text{sys}}\sum_{T\in\mathcal{T}}\frac{A_T}{3}
    \sum_{j=1}^{3}f\left(\bm{q}^{T}_{j}\right)\,.
    \label{eq_2d_quadrature}
\end{equation}
This rule integrates the piecewise-linear interpolation of the sampled integrand $(f)$ exactly. We use the same triangulation $\mathcal{T}$ for the norm, potential-energy density, and gradient-energy density. Sampling determines the spatial resolution, while triangle areas $A_T$ determine the integration weights, no inverse sampling-density factor is used.

\paragraph{Spatial derivatives.}
While the Delaunay triangulation provides stable integration routine, differentiating the corresponding piecewise-linear interpolation is numerically unstable. We therefore decouple
differentiation from the triangulation by evaluating the wave function at additional stencil points around each sampled coordinate $\bm q_i$. For each sampled point $\bm q_i\in \mathcal{T}$, the triangulation $\mathcal{T}$ is used to define a local length scale $h_i$ as half the distance to its nearest neighbor in the Delaunay mesh:
\begin{equation}
    h_i = \min_{j\in\mathcal N(i)} |\bm{q}_i-\bm{q}_j|/2\,
\end{equation}
where $\mathcal N(i)$ denotes the set of vertices connected to $\bm q_i$ by a triangulation edge.

Along each coordinate axis, we evaluate the wave function at five nearby points around each sampled coordinate. For the $r_{2}$-axis,
these evaluations are: $\psi_{\mathrm{2D}}(r_{1,i},r_{2,i}+\delta^{r_{2}}_{i,j})$, for $j=1,\ldots,5$,
where the stencil size $\delta^{r_{2}}_{i,j}$ is determined by the local mesh spacing $h_i$.
We approximate the local wave function by the unique fourth-order polynomial $p_i(r_{2})$ passing through these five values and take its derivative
at the central point:
\begin{equation}
    \partial_{r_{2}}\psi_{\text{2D}}(r_{1,i},r_{2,i}) \approx p_i'(r_{2}=r_{2,i})\,.
\end{equation}
The $r_{1}$ derivative is computed analogously. Near the domain boundary, the stencil points are shifted inward while retaining the same polynomial differentiation procedure.

\subsection{Extension of WKB-inspired radial prior to 2D}
\label{app_2d_wkb}

For the two-dimensional representation in reduced cylindrical coordinates $\{r_\perp,z\}$, we extend the one-dimensional WKB-inspired
envelope of Sec.~\ref{sec_asymptotic}, so that the wave function representation is
\begin{equation}
    \psi_{\mathrm{2D}}(r_\perp,z)
    =
    \psi^{\text{NN}}_{\text{2D}}(r_\perp,z)\,
    \mathcal{W}\!\left(\sqrt{r_\perp^2+z^2}\right)
    \mathcal{W}(r_\perp)\,,
    \label{eq_2D_WKB}
\end{equation}
where $\mathcal{W}(r)\equiv
\psi_{\mathrm{WKB}}(r,\mathcal{E}_{\mathrm{WKB}})$ is the same tempered WKB-inspired envelope used in the one-dimensional formulation~(see Sec.~\ref{sec_asymptotic}). The radial factor
$W(\sqrt{r_\perp^2+z^2})$ imposes the physical short- and long-range
asymptotics, while the additional factor $W(r_\perp)$ suppresses the
wave function near the cylindrical axis $r_\perp\to0$, where the
transformed two-dimensional Hamiltonian contains a singular $1/r_\perp^2$ contribution. The product therefore serves as a regularizing asymptotic prior rather
than as a direct WKB approximation to the full two-dimensional wave function.

For the three-body calculation, we express the WKB-inspired envelope in
terms of the Jacobi radii $r_{1}$ and $r_{2}$. With the equal-mass hyperradius
\(\rho = \sqrt{r_{1}^2 + 4r_{2}^2/3}\), we write the reduced radial amplitude
for each dual-angular channel $\{l,m_l;L,M_L\}$ as
\begin{equation}
    \psi_{l,m_l;L,M_L}(r_{1},r_{2})
    =
    \psi_{l,m_l;L,M_L}^{\mathrm{NN},\text{2D}}(r_{1},r_{2})\,
    \mathcal{W}(r_{1})\,\mathcal{W}(\rho)\,\frac{r_{2}}{\rho}\,.
    \label{eq_3B_WKB}
\end{equation}
The factor $\mathcal{W}(r_{1})$ suppresses
configurations in which the selected pair approaches coalescence, while
$\mathcal{W}(\rho)$ supplies decay at large hyperradius, including
when the third particle recedes at fixed pair separation. At fixed
$r_{1}>0$, the factor $r_{2}/\rho$ makes the envelope vanish linearly as
$r_{2}\to0$. The envelope thus provides a common asymptotic prior for all
channels, while $\psi_{l,m_l;L,M_L}^{\mathrm{NN},\text{2D}}(r_{1},r_{2})$ captures their distinct
radial dependence.

\subsection{Two-dimensional training versus radial-channel training}
\label{app_2d_training}

Both representations minimize the same physical energy functional,
but differ in how much structure is supplied analytically. In the
two-dimensional formulation, the network learns the joint
dependence on $\{r_\perp,z\}$ through the representation~\eqref{eq_2D_WKB},
including the remaining angular structure.
In the radial-channel formulation~\eqref{eq_two_body_representaiton},
the angular dependence is represented analytically by spherical
harmonics. The network therefore learns only the radial channel
functions $\psi^{\text{NN}}_{l,m}(r)$, while angular kinetic and potential couplings are handled
through the centrifugal terms and precomputed Gaunt coefficients.

Removing this decomposition substantially increases the numerical
burden. The 2D samples must resolve both radial localization and
angular variation, while evaluation of the energy additionally
requires Delaunay integration and spatial derivatives along both
$r_\perp$ and $z$. This leads to slower convergence and more expensive
updates than in the radial-channel formulation. At the same time, the
2D calculation provides a controlled test in which the analytical
angular decomposition is removed while retaining the same variational
objective, and separately validates optimization with two continuous
coordinates required by the three-body radial representation.

\section{Extended Results}
\label{app_extended_experiments}

\subsection{Wave function input scale}

\begin{figure}[h]
    \centering
    %\printsizes
    \vspace{-1.em}
    \begin{minipage}[c]{0.59\textwidth}
        \caption{
        \textbf{Logarithmic versus linear discretization.} In linear coordinates (main), a localized peak is poorly resolved due to the large dynamic range, while logarithmic coordinates (inset) distribute resolution across scales and capture the feature more accurately.
        }
        \label{fig_radial_scale}
    \end{minipage}%
    \hfill
    \begin{minipage}[c]{0.4\textwidth}
        \vspace{0.25em}
        \centering
        %\printsizes
    \includegraphics[width=\linewidth]{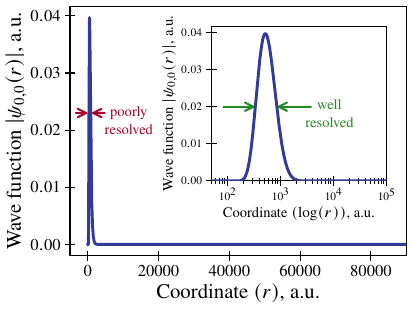}
    \end{minipage}
\end{figure}

\subsection{Qualitative wave function results}

\begin{figure}[H]
% \hfill
    \centering
    %\printsizes
    \includegraphics[width=0.9\textwidth]{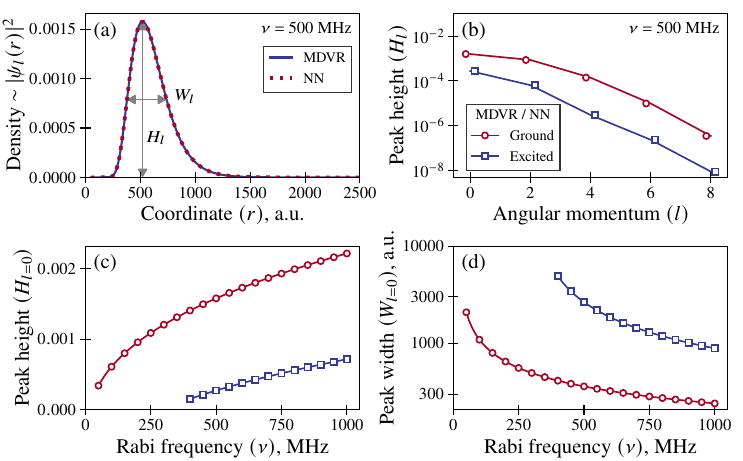}
\caption{
    \textbf{Radial density profiles and peak characteristics of the ground and excited states.} (a) MDVR (solid line) and neural (dotted line) radial densities at $\nu=500\,\text{MHz}$ for $l=0$ channel. The arrows indicate the peak height $H_l$ and full width at half maximum $W_l$. (b) Peak height versus angular momentum $l$ at the same frequency. (c,d) Peak height $H_{l=0}$ and width $W_{l=0}$, respectively, versus Rabi frequency $\nu$. 
    }
    \label{fig_wavefunction_peaks}
\end{figure}

Figure~\ref{fig_wavefunction_peaks}(a) shows that the neural wave function reproduces the MDVR radial density profile at $\nu=500\,\text{MHz}$ ($l=0$ radial channels are presented). We characterize the wave function peak in each angular momentum channel $l$ by its height $H_l$ and its full width at half maximum $W_l$. For both the ground and excited states, the peak heights $H_l$ decrease rapidly with $l$, indicating that higher-order channels contribute less to the wave function (see Fig.~\ref{fig_wavefunction_peaks}(b)). As the Rabi frequency $\nu$ increases, $H_{l=0}$ increases and $W_{l=0}$ decreases for both states (see Fig.~\ref{fig_wavefunction_peaks}(c,d)). The peak of the dominant $l=0$ radial-channel wave function thus becomes sharper, as the depth of the potential well increases $(\sim\nu)$.

\hfill

%\begin{figure}[H]
% \hfill
%    \centering
    %\printsizes
%    \includegraphics[width=0.9\textwidth]{figures/optimized_wavefunction_vs_gt_ground_state.pdf}
%\caption{
%\textbf{Ground-state wave functions across interaction strengths $\nu$.}
%Solid lines denote MDVR reference solutions and dashed lines denote the neural predictions. The model accurately captures both localized and spatially extended regimes across all spherical channels. Larger discrepancies appear primarily in higher-order spherical channels, whose amplitudes are substantially smaller and therefore contribute only weakly to the total energy.
%}
%    \label{fig_qualitative_ground_state}
%\end{figure}

%\begin{figure}[H]
%    \centering
%    \includegraphics[width=0.9\textwidth]{figures/optimized_wavefunction_vs_gt_excited_state.pdf}
%    \caption{
%\textbf{Excited-state wave functions across interaction strengths $\nu$.}
%Solid lines denote MDVR reference solutions and dashed lines denote the neural predictions. The model reproduces the nodal structure and asymptotic decay of the excited states across interaction regimes. Deviations are largest in higher-order spherical channels, which have comparatively small amplitudes and limited impact on the total energy.
%}
%    \label{fig_qualitative_excited_state}
%\end{figure}

\subsection{Ablations}
\label{app_extended_ablation}

\subsubsection{Effect of the WKB envelope spread on the energy and wave function}
\label{app_wkb_spread}

The WKB-inspired envelope, introduced in Sec.~\ref{sec_asymptotic}, is evaluated at the proxy energy $\mathcal{E}_{\mathrm{WKB}}$ with the spread defined as $\kappa_{\mathcal{E}}(r)\to\beta\kappa_{\mathcal{E}}(r)$, with $\beta=0.01$ in the main results. In the main results, the training points are also sampled from this envelope (Sec.~\ref{sec_sampling}), so $\beta$ changes both the form of the wave function and where it is trained, and a stiffer envelope places fewer points at large $r$. To separate the two, we train here on a fixed log-spaced grid of $40000$ points over $r\in[50,10^{5}]$\,a.u., identical for all $\beta$, so that any change in accuracy comes from the envelope itself and not from the placement of the grid points. We repeat the optimization for $\beta\in\{0,\,0.001,\,0.01,\,0.1,\,1\}$ with three seeds each, for the ground state at $\nu\in\{50,100,500,1000\}$\,MHz and for the first excited state at $\nu\in\{400,500,700,1000\}$\,MHz, where it is bound. For the excited state, the ground state of the same $\nu$, $\beta$ and seed is projected out. Here $\beta=0$ removes the decay of the envelope and $\beta=1$ is the literal WKB envelope. We measure accuracy by the relative error~\eqref{eq_relative_error} between the energy at the end of training and the reference MDVR energy.

\begin{figure}[H]
\centering
\includegraphics[width=0.9\linewidth]{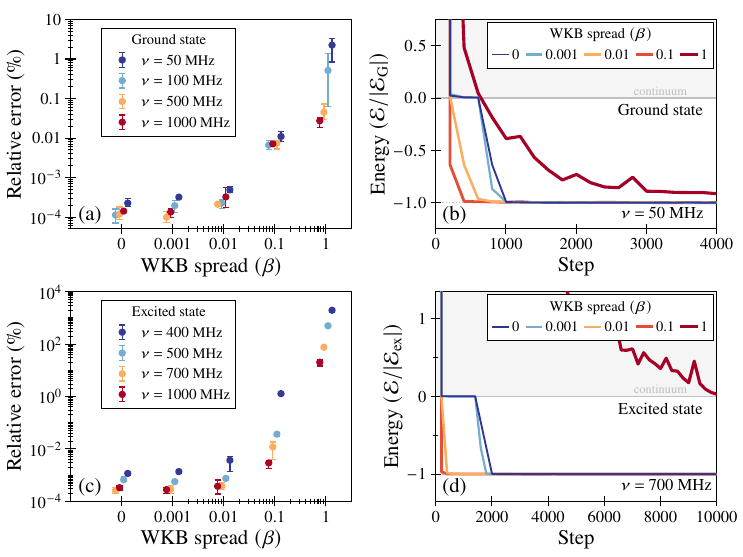}
\caption{\textbf{Effect of the WKB envelope spread $\beta$ on ground and excited state energy and training dynamics}. (a,c) Relative errors of the final ground- and first-excited-state energies, respectively, with respect to MDVR results. Markers show the mean over three random seeds, and whiskers indicate the range. (b,d) Energy during training for the ground state at $\nu=50\,\text{MHz}$ and the first excited state at $\nu=700\,\text{MHz}$, respectively, normalized by the magnitude of the corresponding MDVR energy. The horizontal gray line at $\mathcal{E}=0$ marks the dissociation threshold (continuum edge).}
\label{fig_wkb_sweep}
\end{figure}

For the ground state, the accuracy is insensitive to the spread over two orders of magnitude. All runs with $\beta\le0.1$ agree with MDVR to within about $0.01\%$. Only the literal envelope ($\beta=1$) is noticeably less accurate, and its error grows towards low frequency, reaching about $2\%$ at $\nu=50$\,MHz. At low $\nu$ the proxy energy $\mathcal{E}_{\mathrm{WKB}}$ lies much deeper than the ground-state energy, so the literal envelope decays from a turning point well inside the true wave function. The spread mainly affects how fast the energy converges (Fig.~\ref{fig_wkb_sweep}(b)). A moderate spread ($\beta=0.01$ and $\beta=0.1$) brings the energy to the MDVR value within a few hundred steps, while with absent or very weak envelope decay ($\beta=0$ and $\beta=0.001$) it first stays near the threshold and converges a few times more slowly. For $\beta=1$ the energy approaches the reference slowly from above and stays above it until the end of training.

The excited state is more sensitive to the spread. A weak spread ($\beta\le0.01$) still agrees with MDVR to within about $0.005\%$, but already $\beta=0.1$ loses accuracy at the lowest frequency, and the literal envelope fails at every frequency, with the state ending above the dissociation threshold at low $\nu$. Since the excited state lies much closer to threshold than the ground state, a stiff envelope cuts off most of its tail. The effect on convergence is also stronger (Fig.~\ref{fig_wkb_sweep}(d)). Without a decaying envelope ($\beta=0$ and $0.001$) the energy stays at the threshold for a long time before dropping, a moderate spread ($\beta=0.01$ and $0.1$) brings it to the MDVR value within a few hundred steps, and for $\beta=1$ it is still decreasing at the end of training.

\begin{figure}[H]
\centering
\includegraphics[width=0.9\linewidth]{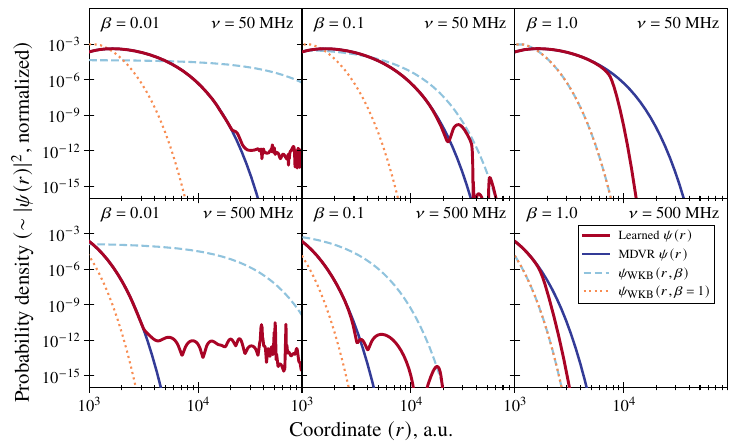}
\caption{\textbf{Effect of the WKB envelope spread $\beta$ on the ground-state wave-function tails.} Rows correspond to $\nu\in\{50,500\}\,\text{MHz}$ and columns to $\beta\in\{0.01,0.1,1\}$. The normalized densities obtained from the learned wave function and MDVR reference are compared to the tempered WKB envelope with the indicated $\beta$ and the untempered envelope with $\beta=1$.}
\label{fig_wkb_tail}
\end{figure}

\begin{figure}[H]
\centering
\includegraphics[width=0.9\linewidth]{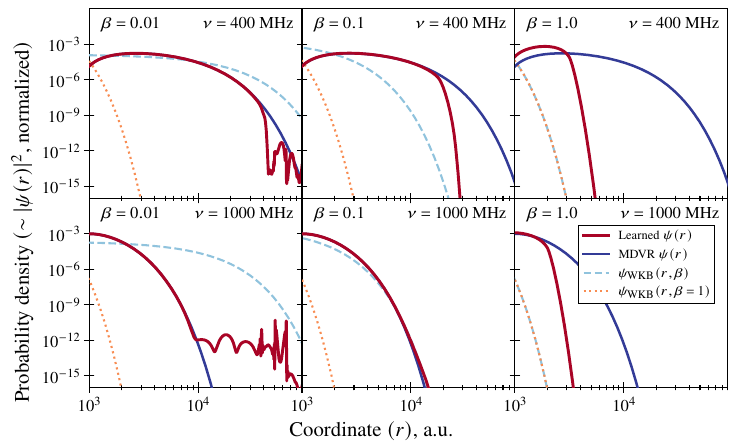}
\caption{\textbf{Effect of the WKB envelope spread $\beta$ on the first-excited-state wave-function tails.} Rows correspond to $\nu\in\{400,1000\}\,\text{MHz}$ and columns to $\beta\in\{0.01,0.1,1\}$. The normalized densities obtained from the learned wave function and MDVR reference are compared to the tempered WKB envelope with the indicated $\beta$ and the untempered envelope with $\beta=1$.}
\label{fig_wkb_tail_excited}
\end{figure}

Figures~\ref{fig_wkb_tail} and \ref{fig_wkb_tail_excited} show the learned tails. For $\beta=0.01$ and $0.1$ the learned density decays well below the envelope and follows the MDVR tail over many orders of magnitude, so the network learns the asymptotic decay rather than inheriting it from the prior. For $\beta=1$ the network cannot undo the overly fast decay of the envelope, and the learned tail falls off faster than MDVR for both states.

\subsubsection{Effect of the angular momentum cutoff}
\label{app_lmax}

\begin{figure}[h]
    \centering
    %\printsizes
    \vspace{-1em}
    \begin{minipage}[c]{0.34\textwidth}
        \caption{\textbf{Angular-momentum cutoff ablation for two-body (top row) and three-body (bottom row) ground states.} (a,c) Training dynamics with different angular cutoffs for $\nu=1000\,\text{MHz}$ and $\nu\approx69\,\text{MHz}$, respectively. Optimization converges even with small cutoffs, but the resulting energies retain substantial angular-truncation errors. (b,d) Converged energies versus $l_{\max}$ and $\{l_{\max},L_{\max}\}$, respectively, at the indicated field strengths. Energies are normalized by the magnitude of the corresponding MDVR (for two bodies) and 2D DVR (for three bodies) reference energies.}
        \label{fig_experiment_three_body_angular}
    \end{minipage}%
    \hfill
    \begin{minipage}[c]{0.65\textwidth}
        \centering
        %\printsizes
    \includegraphics[width=0.99\linewidth]{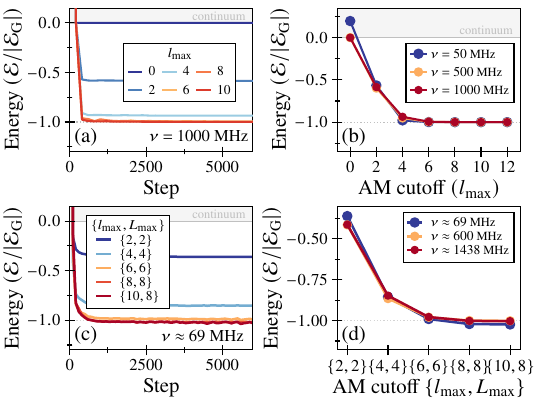}
    \end{minipage}
    \vspace{-1em}
\end{figure}

In the two-body calculation, the spherical-harmonic expansion, introduced in Sec.~\ref{sec_spherical}, is truncated at a finite angular momentum $l_{\max}$. Main text results use the same cutoff as the MDVR reference: $l_{\max}=8$. To check that this truncation does not omit relevant contributions, we repeat the ground-state optimization with $l_{\max}\in\{0,2,4,6,8,10,12\}$ and $\nu\in\{50,100,500,1000\}\,\text{MHz}$ for three random seeds each, with all other settings kept fixed. Figure~\ref{fig_experiment_three_body_angular} shows the training curves at $\nu=1000\,\text{MHz}$, as well as the final energies at three different Rabi frequencies.

With $l_{\max}=0$ no bound state forms, because the dipolar term has a vanishing matrix element within the $l=0$ channel, leaving only the repulsive contribution. With $l_{\max}=2$ the state is bound but the energy is still $40$--$43\%$ above MDVR. Truncating at $l_{\max}=4$ raises the energy by $2$--$6\%$, and $l_{\max}=6$ still leaves an error of $0.04$--$0.5\%$ that grows with $\nu$, reflecting the stronger anisotropic coupling in deeper potentials. From $l_{\max}=8$ onwards the energy is stable, and all errors are below $0.02\%$, comparable to the spread between seeds. The angular expansion is therefore converged at $l_{\max}=8$, and the truncation error is negligible compared to the reported accuracy.

We also test angular convergence in the three-body calculation by increasing the dual-angular cutoff $\{l_{\max},L_{\max}\}$ through $\{2,2\}$, $\{4,4\}$, $\{6,6\}$, $\{8,8\}$, and $\{10,8\}$.
As shown in Fig.~\ref{fig_experiment_three_body_angular}(c,d), optimization converges even at small cutoffs, but the resulting energies retain substantial angular-truncation errors.
Successive basis enlargements produce progressively smaller energy changes.
In particular, the final increase from $\{8,8\}$ to $\{10,8\}$ enlarges the basis from 145 to 190 channels and changes the ground-state energy by at most $0.25\%$ at the tested frequencies.
This weak sensitivity to the final basis enlargement supports the cutoff $\{10,8\}$ used in the three-body benchmarks.

\subsubsection{Effect of sampling strategy}
\label{app_exp_sampling}

\begin{figure}[H]
\centering
\includegraphics[width=0.99\linewidth]{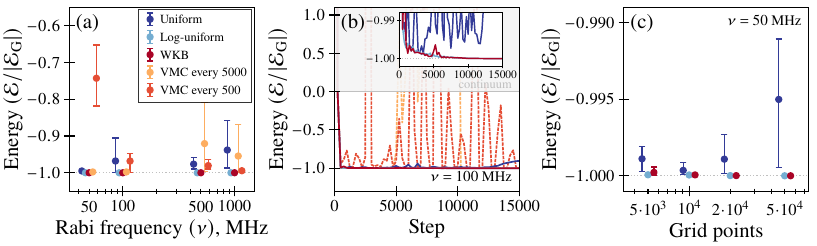}
\caption{\textbf{Effect of fixed and dynamically resampled training grids on ground-state energies.} Energies are normalized by the magnitude of the MDVR ground-state energy. (a) Final energies for fixed uniform, log-uniform, and WKB-guided grids and for grids dynamically resampled from the model every $500$ or $5000$ steps via VMC, each containing $5\cdot10^4$ grid points. (b) Training dynamics for $\nu=100\,\text{MHz}$. Solid lines represent fixed grids and dashed lines represent dynamically resampled grids. The inset magnifies the fixed-grid results near the MDVR energy. (c) Final energies versus the number of grid points for the fixed grids and $\nu=50\,\text{MHz}$. In panels (a,c), markers show the mean over three seeds and whiskers indicate the range. Points are slightly shifted horizontally for visibility.
}
\label{fig_sampling}
\end{figure}

To test whether keeping the coordinate grid fixed throughout the training matters, we also train the network on grids redrawn by VMC every $500$ or $5000$ gradient steps from the evolving model density $|\Psi_{\boldsymbol{w}}(\boldsymbol{r})|^2$, where $\bm{\omega}$ are varying model weights. We keep the number of grid points fixed to $5\cdot10^4$, and train each configuration for $15000$ steps. The final result for the ground-state energy is evaluated on an independent dense grid with $10^5$ points. Figure~\ref{fig_sampling} provides the comparison between (a)~the final energies for different Rabi frequencies $\nu$, (b)~the training dynamics for $\nu=100\,\text{MHz}$, and (c)~the final energies for different numbers of grid points in the evaluation grid.

The fixed WKB-guided and log-uniform grids reach relative errors below $0.015\%$ with respect to the MDVR, and settle within $600$--$2100$ steps, while the uniform grid stays $0.5$--$6\%$ above the reference. Resampling from the model via VMC is both less accurate and less stable, with errors of $0.2$--$8\%$, when redrawn every $5000$ steps, and up to $26\%$, when redrawn every $500$ steps. As Fig.~\ref{fig_sampling}(b) indicates, each grid re-sampling changes the discrete objective and can push the energy into the continuum, so the optimization has to repeatedly recover. The fixed WKB-guided and log-uniform grids remain within $0.03\%$ of the MDVR reference down to $5\cdot10^3$ evaluation grid points, whereas the uniform grid stays $0.04$--$0.5\%$ above it at all sizes. Having a fixed non-uniform grid is therefore essential for stable and accurate optimization.

% (a) Final ground-state energy for grids of $5\times10^4$ points. Solid lines are fixed grids, and dashed lines are grids resampled from the model every $5{,}000$ or $500$ steps. (b) Energy during training at $\nu=100$\,MHz for one seed. The inset magnifies the fixed grids near the MDVR energy. (c) Final ground-state energy of the fixed grids versus the number of grid points at $\nu=50$\,MHz. In (a) and (c), markers show the mean over 3 seeds, whiskers the range from minimum to maximum, and points are shifted horizontally for visibility. One diverged run (uniform grid, $5\times10^3$ points) is excluded.

\subsubsection{Independent evaluation of trained networks}
\label{app_evaluation}

\begin{figure}[h]
    \centering
    %\printsizes
    \vspace{-1em}
    \begin{minipage}[c]{0.58\textwidth}
        \caption{\textbf{Independent evaluation of fixed trained networks.} Relative error with respect to MDVR, in percent, of the energy evaluated on the training grid and on independent logarithmic and uniform grids. Markers show the mean over 3 seeds and whiskers the range from minimum to maximum. Points are shifted horizontally for visibility.}
        \label{fig_evaluation}
    \end{minipage}%
    \hfill
    \begin{minipage}[c]{0.4\textwidth}
        \centering
        %\printsizes
    \includegraphics[width=\linewidth]{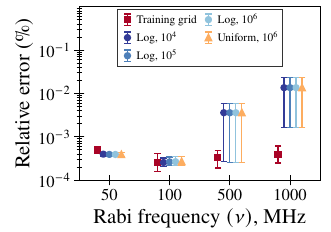}
    \end{minipage}
    \vspace{-1em}
\end{figure}

During training, the energy is estimated by a quadrature on the training points, for which the network is optimized, so this estimate can be biased. To validate the accuracy of our neural representation, we hold the trained ground-state networks fixed for $\nu\in\{50,100,500,1000\}\,\text{MHz}$ and three random seeds, and re-evaluate their energy estimate on independent logarithmic grids with $10^4$, $10^5$ and $10^6$ grid points and a uniform grid with $10^6$ grid points. Figure~\ref{fig_evaluation} displays the relative error~\eqref{eq_relative_error} between the energy evaluated on an independent grid and the MDVR reference.

All independent evaluations agree to within $5\cdot10^{-5}\%$ and lie above MDVR, as the variational principle suggests. The training-grid estimate is up to $0.023\%$ lower and hides the growth of the error with $\nu$. Energies below the reference therefore come from evaluating on the training points, where the network exploits quadrature errors, together with mini-batch fluctuations and an envelope evaluated on those same points. 

% \subsection{Effect of sampling strategy}
% \label{app_exp_sampling}

% We compare three sampling strategies: uniform, logarithmic, and the proposed WKB-based MCMC sampling. Uniform sampling leads to slow convergence and higher final energy due to poor coverage of relevant regions (Fig.~\ref{fig_grid}). Both logarithmic and WKB-based sampling improve optimization.
% While the logarithmic grid performs comparably in this setting, it relies on a fixed, problem-specific coordinate transformation. In contrast, the proposed approach adapts to the physical structure through the sampling distribution itself and does not require manual tuning. This distinction becomes critical in higher-dimensional settings, where suitable coordinate transformations are unavailable.

% \begin{figure}[H]
%     \centering
%     \includegraphics[width=0.6\textwidth]{figures/grid_comparison.pdf}
% \caption{
% \textbf{Effect of sampling strategy.} Training trajectories for ground and excited states using uniform, logarithmic, and WKB-based sampling. Uniform sampling leads to noisy optimization and slow convergence, while logarithmic and WKB-based sampling improve stability and convergence.}
%     \label{fig_grid}
% \end{figure}

\subsection{Robustness across random seeds}
\label{app_exp_robustness}
% \begin{figure}[H]
%     \centering
%     \includegraphics[width=0.9\textwidth]{figures/energy_across_seeds_by_nu.pdf}
%     \caption{
% \textbf{Optimization trajectories across random seeds.}
% Training curves for different random initializations across interaction regimes. The model consistently converges to similar final energies, indicating stable optimization behavior despite stochastic initialization.
% }
%     \label{fig_seeds_ablation}
% \end{figure}

To evaluate optimization stability and reproducibility, we repeated training using five independent random seeds across representative field strengths ($\nu$, sampled in $100\,\text{MHz}$ increments). Table~\ref{table_error_bars} reports the resulting mean energies and standard deviations alongside baseline benchmark values computed via mapped discrete variable representation (MDVR). While absolute variability increases at larger $\nu$, energy magnitudes scale proportionally, keeping the relative variance small across all regimes. Overall, the optimization remains stable across random initializations and all physical trends are preserved.

%\begin{table}[htbp]
%\centering
%\caption{Ground-state and excited-state energies across representative interaction regimes ($\nu$), reported as mean $\pm$ standard deviation across 5 random seeds compared to MDVR benchmarks.}
%\label{table_error_bars}
%\small
%\begin{tabular}{r ccc ccc}
%\toprule
%& \multicolumn{2}{c}{\textbf{Ground State (kHz)}} & \phantom{a} & \multicolumn{2}{c}{\textbf{Excited State (kHz)}} \\
%\cmidrule{2-3} \cmidrule{5-6}
%$\nu$ (MHz) & Ours & MDVR && Ours & MDVR \\
%\midrule
%100  &   $-57.963 \pm 0.030$  &   $-57.948$ &&         --         &     --    \\
%200  &  $-215.458 \pm 0.101$  &  $-215.223$ &&         --         &     --    \\
%300  &  $-425.085 \pm 0.589$  &  $-425.145$ &&         --         &     --    \\
%400  &  $-670.347 \pm 1.169$  &  $-670.824$ &&   $-1.005 \pm 0.145$ &   $-1.214$ \\
%500  &  $-944.922 \pm 1.340$  &  $-943.628$ &&   $-5.397 \pm 0.562$ &   $-5.757$ \\
%600  & $-1237.390 \pm 4.307$  & $-1238.276$ &&  $-12.386 \pm 2.985$ &  $-14.252$ \\
%700  & $-1545.055 \pm 16.572$ & $-1551.201$ &&  $-27.791 \pm 0.564$ &  $-26.914$ \\
%800  & $-1883.576 \pm 6.018$  & $-1879.825$ &&  $-44.475 \pm 1.020$ &  $-43.781$ \\
%900  & $-2227.304 \pm 9.927$  & $-2222.201$ &&  $-65.399 \pm 2.503$ &  $-64.806$ \\
%1000 & $-2577.781 \pm 41.851$ & $-2576.802$ &&  $-90.684 \pm 3.324$ &  $-89.896$ \\
%\bottomrule
%\end{tabular}
%\end{table}

\begin{table}[htbp]
\centering
\caption{Ground-state and excited-state energies across representative interaction regimes ($\nu$), reported as mean $\pm$ standard deviation across 5 random seeds compared to MDVR benchmarks.}
\label{table_error_bars}
\small
\begin{tabular}{r cc @{\qquad} cc}
\toprule
& \multicolumn{2}{c}{\textbf{Ground State (kHz)}}
& \multicolumn{2}{c}{\textbf{Excited State (kHz)}} \\
\cmidrule(lr){2-3} \cmidrule(lr){4-5}
$\nu$ (MHz) & Ours & MDVR & Ours & MDVR \\
\midrule
100  & $-57.94733 \pm 0.00006$  & $-57.948$   & --                       & --        \\
200  & $-215.22267 \pm 0.00020$ & $-215.223$  & --                       & --        \\
300  & $-425.14355 \pm 0.00023$ & $-425.145$  & --                       & --        \\
400  & $-670.805 \pm 0.025$     & $-670.824$  & $-1.21359 \pm 0.00005$   & $-1.214$  \\
500  & $-943.601 \pm 0.024$     & $-943.628$  & $-5.75719 \pm 0.00038$   & $-5.757$  \\
600  & $-1238.263 \pm 0.010$    & $-1238.276$ & $-14.25189 \pm 0.00010$  & $-14.252$ \\
700  & $-1551.183 \pm 0.012$    & $-1551.201$ & $-26.91368 \pm 0.00020$  & $-26.914$ \\
800  & $-1879.707 \pm 0.215$    & $-1879.825$ & $-43.778 \pm 0.005$      & $-43.781$ \\
900  & $-2222.145 \pm 0.069$    & $-2222.201$ & $-64.803 \pm 0.003$      & $-64.806$ \\
1000 & $-2576.533 \pm 0.237$    & $-2576.802$ & $-89.888 \pm 0.007$      & $-89.896$ \\
\bottomrule
\end{tabular}
\end{table}

\subsection{Additional classical baseline}
\label{app_exp_weinberg}

The MDVR method is based on the discrete variable representation (DVR), also known as the pseudospectral method. For smooth wave functions, the discretization error of pseudospectral methods decreases exponentially as the number of grid points increases. MDVR further improves the efficiency of DVR by introducing an optimized coordinate mapping, allowing the same level of accuracy to be achieved with significantly fewer grid points. Consequently, for the present problem, the numerical error introduced by the MDVR method can be made negligibly small by using a sufficiently dense grid.

To validate the accuracy of the MDVR approach, we compare its results with those obtained from an independent momentum-space calculation based on the Weinberg expansion \citep{Weinberg1963Quasiparticles} of the interaction potential. As shown in Table 2, the two methods agree to within $0.02\%$ for the ground-state energy, demonstrating the high accuracy of the MDVR method. The relative error first decreases and then gradually increases with increasing energy, reflecting the competing effects of the increasing absolute deviation (numerator) and the increasing binding energy (denominator). The absolute deviation may arise from numerical errors in both methods.

\begin{table}[H]
\centering
\caption{Comparison of MDVR with an independent Weinberg-expansion (WeinE) benchmark}
\label{tab:mdvr_weine_comp}
\begin{tabular}{r r r r}
\toprule
$\nu$ (MHz) & WeinE energy (kHz) & MDVR energy (kHz) & Relative error \\
\midrule
50   &     $-11.071$ &     $-11.069$ & $0.018\%$   \\
100  &     $-57.952$ &     $-57.948$ & $0.007\%$   \\
200  &    $-215.227$ &    $-215.223$ & $0.002\%$   \\
300  &    $-425.143$ &    $-425.145$ & $<0.001\%$  \\
400  &    $-670.813$ &    $-670.824$ & $0.002\%$   \\
500  &    $-943.603$ &    $-943.628$ & $0.003\%$   \\
600  &   $-1238.235$ &   $-1238.276$ & $0.003\%$   \\
700  &   $-1551.141$ &   $-1551.201$ & $0.004\%$   \\
800  &   $-1879.743$ &   $-1879.825$ & $0.004\%$   \\
900  &   $-2222.096$ &   $-2222.201$ & $0.005\%$   \\
1000 &   $-2576.673$ &   $-2576.802$ & $0.005\%$   \\
\bottomrule
\end{tabular}
\end{table}

\section{Technical Details}
\label{app_technical_details}

\paragraph{Optimization.}
All experiments were implemented in PyTorch and optimized using Adam with learning rate $10^{-3}$, gradient clipping threshold $10^{-3}$, and a cosine learning-rate scheduler decaying to $10^{-5}$. Models were trained for $3\times10^3$ epochs ($\approx1.2\times10^4$ steps), although convergence typically occurred substantially earlier. All computations were performed in double precision. Experiments were run on a single NVIDIA H100 GPU. The radial-channel experiments used a batch size of $10^4$, with training time of about $2$ minutes per run.

\paragraph{2D optimization.}
The optimization for the 2D wave function used the same hyperparameters as for the radial case. One noticeable exception is the batch size. In order to accurately represent integrals over 2D coordinate domain, batches of $3\times10^5$ coordinate points were used ($\times30$ compared to the radial case). This substantially increases every step's runtime and GPU memory usage. Nevertheless, the training converges within the aforementioned $3\times10^4$ steps.

\paragraph{Wave function representation.}
The wave function was parameterized using a multilayer perceptron (MLP) with:
\begin{itemize}[leftmargin=2.em]
    \setlength\itemsep{0em}
    \item $6$ hidden layers,
    \item hidden dimension $128$,
    \item sinusoidal positional encoding with $N=3$ frequency bands.
\end{itemize}
These hyperparameters were selected by first training candidate architectures in a supervised setting against reference wave function data, and subsequently using the same architecture in the variational formulation proposed in this work.

The positional encoding was defined as
\[
\mathrm{PE}(r)=
\big[
\sin(2^k 2\pi \tilde r),
\cos(2^k 2\pi \tilde r)
\big]_{k=0}^{N-1},
\]
where $\tilde r = (r-r_{\min})/(r_{\max}-r_{\min})$ denotes the normalized radial coordinate.

The model outputs correspond to even spherical harmonic channels up to $l_{\max}=8$, namely $l=0,2,4,6,8$. The WKB-inspired asymptotic envelope described in Sec.~\ref{sec_asymptotic} was tempered using $\beta=0.01$ to avoid over-constraining the representation.

\paragraph{Sampling grids.}
For the sampling-grid comparison experiments, each grid contained $5\times10^4$ points over the radial domain $r \in [50,10^5]$. The uniform grid used uniformly random points, while the logarithmic grid used log-uniformly random points over the same interval. The WKB-based grid was constructed using the MCMC sampling procedure described in Sec.~\ref{sec_sampling}.

\paragraph{VMC sampling.}
The VMC sampler used $5000$ walkers in a one-dimensional radial space with cutoff range $r \in [50,10^5]$. Sampling was performed in log-space to efficiently resolve the multiple spatial scales present in the problem. A symmetric Gaussian proposal with initial step size $0.5$, burn-in length $5000$, and thinning factor $10$ was used. After burn-in, $10$ samples were retained per walker, resulting in a total of $5\times10^4$ sampled points.

During burn-in, the proposal step size was adaptively adjusted using a windowed acceptance-rate controller targeting an acceptance probability of $0.1$. Let $a_t$ denote the acceptance rate measured over an adaptation window of size $100$, and let $a^\star$ denote the target acceptance rate. The proposal scale was updated according to
\[
\log s_{t+1}
=
\log s_t
+
\eta \,
\mathrm{clip}(a_t-a^\star,-c,c),
\]
where $s_t$ is the proposal step size, $\eta=0.6$ is the controller learning rate, and $c=0.2$ is a clipping threshold. The multiplicative update was additionally constrained using growth and shrinkage caps of $1.6$ per adaptation step to prevent unstable changes in proposal variance.

\paragraph{Energy scaling.}
We work within the atomic unit system. In these units, typical energy values for our system are on the order of $10^{-12}$ (Hartrees). Such extremely small numerical values lead to poor numerical conditioning during optimization. To improve stability, we rescale the energies using the reference value $|\mathcal{E}_{\text{WKB}}|$, defined in Sec.~\ref{sec_asymptotic}. All reported energies were normalized by this reference scale during training and evaluation.

\subsection{Physical Hyperparameters}

\textbf{Choice of $\nu$.} 
The first and second bound states appear when $\nu > 16\,\text{MHz}$ and $\nu > 300\,\text{MHz}$, respectively. We choose our benchmark dataset within $\nu \in [50, 1000]\,\text{MHz}$ for the ground state and $\nu \in [400, 1000]\,\text{MHz}$ for the excited state, though our approach applies across other $\nu$ ranges.

\textbf{Choice of $r$.} 
The interaction potential features a repulsive barrier around $700\,\text{a.u.}$ (decreasing slightly with increasing $\nu$), which exponentially suppresses the wave function amplitude at smaller distances. Consequently, at $r_{\min} = 50\,\text{a.u.}$, the wave function amplitude already approaches zero, allowing us to exclude the origin and small $r$ values that would otherwise induce numerical instability. Similarly, setting $r_{\max} = 10^5\,\text{a.u.}$ ensures sufficient wave function decay at large distances.

\end{document}

%% file: math_commands.tex
\usepackage{amsmath,amsfonts,bm}

\def\eqref#1{equation~\ref{#1}}
\def\1{\bm{1}}

\DeclareMathAlphabet{\mathsfit}{\encodingdefault}{\sfdefault}{m}{sl}
\SetMathAlphabet{\mathsfit}{bold}{\encodingdefault}{\sfdefault}{bx}{n}